\documentclass[aps,pre,twocolumn,float,floatfix,epsfig,pdflatex,notitlepage]{revtex4-1}
\usepackage{color,ulem,verbatim,bbold,float}
\usepackage{amssymb,amsbsy,amsmath,mathrsfs}
\usepackage{times}
\usepackage{amssymb}
\usepackage{amsbsy}
\usepackage{amsmath}
\usepackage{float}
\usepackage{graphicx}
\usepackage{color}
\usepackage{enumitem}
\usepackage{bm}
\usepackage[utf8]{inputenc}
\usepackage{mathtools}
\usepackage[plainpages=false,pdfpagelabels,colorlinks=true,linkcolor=red,urlcolor=blue,citecolor=blue,pdftitle={},pdfauthor={},pdfdisplaydoctitle={},pdfduplex=DuplexFlipLongEdge]{hyperref}
\usepackage{natbib}
\newcommand\bea{\begin{eqnarray}}
\newcommand\eea{\end{eqnarray}}
\newcommand\beq{\begin{equation}}
\newcommand\eeq{\end{equation}}

\newcommand{\noi}{\noindent}
\newcommand{\non}{\nonumber}
\newcommand{\al}{\alpha}
\newcommand{\de}{\delta}

\newcommand{\ep}{\varepsilon}
\newcommand{\ka}{\kappa}
\newcommand{\lm}{\lambda}
\newcommand{\si}{\sigma}

\newcommand{\om}{\omega}

\newcommand{\pa}{\partial}
\newcommand{\la}{\langle}
\newcommand{\ra}{\rangle}

\newcommand{\Ob}{\mathcal{O}}
\newcommand{\Ik}{\mathcal{I}^{k}}

\newcommand{\I}{\mathcal{I}}
\newcommand{\hd}{h_{D}^{x}}
\newcommand{\Jxij}{J^{x}_{ij}}
\newcommand{\Jyij}{J^{y}_{ij}}
\newcommand{\Jzij}{J^{z}_{ij}}
\newcommand{\Sgn}[0]{\mathrm{Sgn}}
\newcommand{\bra}[1]{\langle #1 |}
\newcommand{\ket}[1]{| #1 \rangle}
\newcommand{\heff}{H_{eff}}

\newcommand{\eps}{\epsilon}

\newcommand{\tc}{\tilde{a}}

\newcommand{\hb}{\hat{b}}
\newcommand{\hbd}{\hat{b}^\dagger}
\newcommand{\hc}{\hat{c}}
\newcommand{\ha}{\hat{a}}

\newcommand{\cN}{\mathcal{N}}
\newcommand{\cM}{\mathcal{M}}

\begin{document}

\title{Statistical Mechanics of Floquet Quantum Matter: Exact and Emergent Conservation Laws}

\author{Asmi Haldar$^{1}$, and Arnab Das$^2$}

\affiliation{$^1$Max Planck Institute for the Physics of Complex Systems, N\"{o}thnitzer Stra{\ss}e 38, 01187 Dresden, Germany}
\affiliation{$^2$Indian Association for the Cultivation of Science (School of Physical Sciences), 2A \& 2B Raja S. C. Mullick Road, Kolkata 700032, India} 

\begin{abstract}
	Equilibrium statistical mechanics rests on the assumption of ergodic dynamics of a system modulo the conservation laws of local observables: extremization of entropy immediately gives Gibbs' ensemble (GE) for energy conserving systems and a generalized version of it (GGE) when the number of local conserved quantities (LCQ) is more than one. Through the last decade, statistical mechanics has been extended to describe the late-time behaviour of
	periodically driven (Floquet) quantum matter starting from a generic state. The structure built on the fundamental assumptions of ergodicity and identification of the relevant ``conservation laws" in this inherently non-equilibrium setting. More recently, it has been shown that the statistical mechanics has a much richer structure due to the existence of {\it emergent} conservation laws: these are approximate but stable conservation laws arising {\it due to the drive}, and are not present in the undriven system. Extensive numerical and analytical results support perpetual stability of these emergent (though approximate) conservation laws, probably even in the thermodynamic limit. This banks on the recent finding of a sharp ergodicity threshold for Floquet thermalization in clean, interacting non-integrable Floquet systems. This opens up a new possibility of stable Floquet engineering in such
systems.  This review intends to give a theoretical overview of 
these developments. We conclude by briefly surveying 
the current experimental scenario. 
\end{abstract}

\maketitle

\section{Introduction}
This is an introductory review of the statistical mechanics of periodically driven 
closed quantum many-body systems, or quantum Floquet matter, developed through the last decade. By `close' we mean it undergoes a pure Schr\"{o}dinger evolution under a time-periodic Hamiltonian, hence its energy is not conserved. The statistical mechanics is for predicting the distribution of a 
subsystem of the entire closed system (or, alternatively, describing the local operators). 
The strategy of building a statistical mechanical description of the steady state of a complex many-body system consists of constructing the most unbiased distribution for the system respecting all the existing constraints. This requires identification of the relevant entropy functional, and then maximizing it subject to the constraints, which, for the purpose of a local description of a system, are the additive conservation laws. Hence the conservation laws are central to the approach described here.\\

The subject of Floquet physics has been existing since decades in various forms in various fields like condensed matter, quantum optics and atomic physics. 
However, the statistical mechanics of periodically driven quantum matter took shape
only over the last decade. Within this period, the research area of Floquet quantum matter
exploded into a voluminous field. Reviewing every aspect of it is far from what this article aims, rather, the focus is on a very specific line of development -- the statistical mechanics describing the late-time behaviour of Floquet quantum matter. In the question of organization, there are several finer classifications of the Floquet phases of matter (based on symmetry classes and other criteria), but in this overview we will focus on a broader classification based on the stability (against unbounded heating) and the underlying conservation laws responsible for it. A summary
of the view presented here
is given in Fig.~\ref{Classification}. \\

\section{Equilibrium Statistical Mechanics and Thermalization: A short recapitulation via Information Theory}
\label{Sec:Eqlb_Stat_Mech}
Statistical 
mechanics is a framework for describing the equilibrium states of matter, while Floquet matter is inherently out of equilibrium. 
Before making a concrete connection between the two apparently disparate subjects, in this section we recall some selected basics of equilibrium statistical mechanics and thermalization which appear useful in the context of Floquet matter.
We resort to the information theoretic approach~\cite{Jaynes}, since it provides
the succinctness just right for our purpose.
\subsection{Unbiased Statistical Distributions and Entropy}
Suppose we know nothing about a variable $x$, and we are asked
to guess the normalized probability distribution $\{P_{i}\}: P_{i} = P(x_{i}),$
\beq
\sum_{i}^{N}P_{i} = 1, 
\label{Norm}
\eeq
over a countable set of values $x_{i}, i=1,N,$ $x$ can take. The most unbiased answer
will obviously be $P_{i} = 1/N$ -- a flat distribution.
We arrive at this conclusion by maximizing the uncertainty -- since we have no information at all, our distribution should reflect that, and hence should be the one with most uncertainty/ignorance. 
Now suppose we have an additional constraint -- we know that whatever the distribution $\{P_i\}$ is, it must give a preassigned value for the expectation 
of a particular quantity, 
$\I,$ say, i.e., $P_i$ must satisfy
\beq
\sum_{i=1}^{N} P_{i}~\I(x_{i}) = {\rm some ~ given ~ value.}
\label{constraint}
\eeq
\noindent
Then what is the unbiased distribution in presence of this constraint?
The answer will now be a distribution that maximizes the uncertainty,
obeying the constraint (since we have no information about the distribution other than that the constrained is satisfied).
In order to maximize uncertainty, we first need a sensible
measure of uncertainty of the probability distribution. This was
provided by Shannon~\cite{Shannon_Entropy_Original} via
his entropy functional 
\beq
S_{s}[\{P_{i}\}] = -K\sum_{i}P_{i}\ln{P_{i}},
\label{Shanon_Entropy}
\eeq
\noi where $K$ is a positive constant. $S_{s}$ (known as Shannon entropy) was shown to have the following properties~\cite{Shannon_Entropy_Original}.
{\bf (a)} It is a faithful measure of the uncertainty in $\{P_{i}\}$ -- it increases monotonically with the uncertainty in $\{P_{i}\}.$  
{\bf (b)} It is always positive. 
{\bf (c)} It is additive: If we consider two separate non-interacting systems, then the total uncertainty of those two taken together should simply add up, and so do their individual $S_{s}$s to give the 
total $S_{s}$
for the systems taken together. 
It was also shown that the form of the functional $S_s$ given in Eq.~\ref{Shanon_Entropy} is the {\it unique} one to satisfy {\bf (a) - (c)}. 
\\
Accepting Shannon entropy as {\it the}  measure of uncertainty in $\{P_i\},$ all we have to do is to maximize (extremization is usually sufficient) $S_{s}$ with respect to $\{P_i\}$, modulo the constraint. This is a simple extremization problem, and can be solved using the standard method of Lagrange's multiplier taking into account the two constraints in Eqs.~(\ref{Norm}) and (\ref{constraint}), employing the corresponding Lagrange multipliers $\lambda_{0}$ and $\lambda_{1}$ respectively. 
Extremizing
\beq
F[\{P_{i}\}] = S_{s}[\{P_{i}\}] + \lm_{0}\sum_{i=1}^{N}P_{i} 
+ \lm_{1}\sum_{i=1}^{N}P_i \I(x_{i})
\eeq
\noi
with respect to $P_{i}$s immediately gives the desired distribution
\beq
P_{i} = e^{-\lambda_{0}}e^{-\lm_{1}{\cal I}(x_i)},
\label{GGE_1}
\eeq
\noi where $Z = e^{\lambda_{0}} = \sum_{i=1,N}e^{\lm_{1}{\cal I}(x_i)}$ is the normalization factor, also known as the partition function. \\

Above can be applied to deduce the unbiased statistical distribution for a sub-system of a large physical system. The meaningful conserved quantities are the {\it local} conserved quantities (LCQ), represented by local operators. 
An operator is said
 to be $k-$local on a lattice if its real space representation is a product of $k$ onsite operators sitting on consecutive sites. We call a $k-$local operator non-local if $k$ is not a fixed number, but scales with the system-size. Intuitively, 
 a $k-$local operator contains information only over
 a finite length scale ($k-$ consecutive sites) on the lattice.
A $k-$local conserved quantity is additive for two large sub-subsystems of linear dimensions
$ \gg k.$ 
Such additive integrals of motion for the entire system (e.g. energy, particle number etc as applicable) are the quantities whose
sharp average values characterize the distribution for the sub-system (see, e.g., ~\cite{Pathria}), and hence play the role of $\I(x)$ for it (here $x$ denotes the complete set of variables/observables necessary for defining the state of a sub-system). If energy is the only conserved quantity of interest, then Eq.~\ref{GGE_1} 
immediately gives the Gibbs's distribution 
\beq
P_{i} = e^{-\beta H(x_i)}/Z,
\label{Gibbs}
\eeq
\noi
where $\I = H$ is the energy (Hamiltonian), $\lm_{1} = \beta = 1/k_{B}\Theta,$ $k_{B}$ being the Boltzmann constant (which we will set to unity henceforth), and $\Theta$ being
the temperature.\\

If there are $L$ number of LCQs denoted by $\Ik$s, 
then the distribution is given by 
\beq
P_{i} = e^{-\sum_{k=1}^{L}\lambda_{k}\Ik}/Z,
\label{GGE_L}
\eeq
\noi and the ensuing ensemble is called the generalized Gibbs' ensemble (GGE). In quantum mechanics, integrability is an ill defined concept, and hence 
there is no concrete prescription to identify the conserved quantities that are relevant for constraining the distribution of the sub-system, except that additive LCQs are always bona fide
candidates. An ab-initio 
numerical verification of GGE was first reported in~\cite{Rigol_GGE_1} (see also~\cite{Rigol_GGE_2}). \\

The statistical entropy of the sub-system is the classical uncertainty in the state of the sub-system (i.e., how mixed it is). This is quantified by the Von Neumann entropy or entanglement entropy (the latter follows from the fact that the entanglement between the sub-system and the rest of the system determines how mixed the state of the sub-system is -- when the entanglement is zero, the sub-system is in a pure state). The Von Neumann entropy is given by
\beq 
S_{E} = - Tr\left[{\rho_{sub}\ln{(\rho_{sub})}}\right]  
= -\sum_{\al=1}^{N_{_D}}P_{\al}\ln{P_{\al}},  
\label{SvN}
\eeq
\noi where $\rho_{sub}$ is the density matrix of the sub-system, $N_{_D}$ is
the total number of states of the sub-system, and $P_{\al}$ is the 
eigenvalues of $\rho_{sub}$ corresponding to the eigenvector $|P_{\al}\ra$,
and gives the probability of obtaining the sub-system in the state $|P_{\al}\ra$ in a projective measurement done in the 
eigen-basis of $\rho_{sub}.$
The final expression of $S_E$ in Eq.~\ref{SvN} is exactly in the form of Shannon
entropy (Eq.~\ref{Shanon_Entropy}), and gives the measure of how mixed the state $\rho_{sub}$ is: if it is a pure state, then $P_{\al} = 0$ for all $\alpha$ except one, for which it is 1. then we get $S_{E} = 0.$ Thus, $S_{E}$
is basically the Shannon entropy characterizing the classical uncertainty 
in the state of the sub-system due to interaction with the
rest of the system. 

\subsection{Thermalization and the Eigenstate Thermalization Hypothesis}
Thermalization broadly means emergence of the thermal behaviour of a sub-system due to interaction with the rest of the system, in a generic many-body system (which, when kept isolated, has energy as the only LCQ). Usually it refers to a dynamical process, where the entire system is allowed to evolve from a non-thermal state, and the chaotic dynamics eventually allows a sub-system to explore its phase space within a narrow energy window fixed by the initial condition. After the steady state is reached, only thing we can guess about a sub-system is that it has a sharply defined average energy determined by the initial state, and following the deduction of Eq.~\ref{Gibbs}, its statics will be given by a Gibbs' distribution.\\

The last two decade have seen the development of a more
stringent concept of thermalization -- it hypothesizes thermal behaviour
of a sub-system right at the level of the eigenstates of a closed quantum chaotic system. This Eigenstate thermalization hypothesis (ETH)~\cite{Srednicki_ETH, Rigol_Nature} says, for a generic interacting many-body quantum system, which has no other LCQ than energy, each eigenstate is thermalized to a temperature compatible with its energy density. In the strongest form, this amounts to saying that 
when the whole system is in any of its eigenstates, the expectation values
of all local observables over the eigenstate equal to their respective 
thermal average value corresponding to the temperature compatible with
the energy density of the eigenstate. In the parlance of information theory discussed above, this means that even being at an eigenstate, the system
provides enough fluctuations to its {\it finite} sub-systems -- sufficient to render its state completely uncertain except fixing its energy. If there are more than 
one conserved quantities that affects the statistical distribution of the
sub-system, then we likewise get a GGE (Eq.~\ref{GGE_L}) in the eigenstate level. At its extreme lies the integrable systems, where there are as many relevant LCQs as there are degrees of freedoms, i.e.,
one has $L=N.$ \\

Even at the level of ETH, the distributions are entirely determined based on (a) the assumption of ergodicity of a sort (albeit in the eigenstate level) and (b) the LCQs. This allows one to predict the long-time average behaviour of the local observables/sub-systems of a closed quantum chaotic system evolving from a typical initial state (a quantum quench), based only on the knowledge of the relevant conserved quantities as we discuss next.

\subsection{Quantum Quenches, Diagonal Ensemble Averages (DEA) and Thermalization}
We now focus on the late time behaviour of the state of a generic system that satisfies ETH. 
It can be argues that if $H$ is the Hamiltonian, and 
$|\psi(t=0)\ra = \sum_{\al}c_{\al}|\ep_{\al}\ra$ is the initial state, where
$|\ep_{\al}\ra$ are the eigenstates of $H$ corresponding to the eigenvalues (quasi-energies) $\ep_{\al},$ then at a later time $t>0$ the expectation value of a local observable $\Ob$ will be
\beq
\langle\psi(t)|\Ob|\psi(t)\ra 
= 
\sum_{\al,\al^{'}}C_{\al}^{*}C_{\al^{'}}e^{-i(\ep_{\al^{'}} - \ep_{\al})t}
\la \ep_{\al}|\Ob|\ep_{\al^{'}}\ra
\label{O_t}
\eeq
The question is, whether it can approach an approximate steady value (with small fluctuations around it) at late times. Here, by late time we mean the time 
longer than that required for the transients to die down and the steady state to appear. Typically a time much longer than the inverse of the relevant spectral gaps (in proper units) qualifies. 
If a steady state is indeed reached, then
this steady value must be equal to
\beq
\la \Ob(t\to\infty)\ra \approx \Ob_{_{DE}} = \sum_{\al}|C_{\al}|^{2}\la \ep_{\al}|\Ob|\ep_{\al}\ra,
\label{DE_O}
\eeq
\noi which is obtained by dropping the explicitly time-dependent terms, i.e., by
retaining only the $\al = \al^{'}$ terms in the above summation~\cite{Rigol_Nature}. 
The expression in Eq.~\ref{DE_O} is exactly identical with the average of the $\Ob$ obtained over
the density matrix
\beq
\rho_{_{DE}} = \sum_{\al}|C_{\al}|^{2}|\ep_{\al}\ra\la \ep_{\al}|,
\label{DE_rho}
\eeq
\noi which is diagonal in the eigen-basis $\{|\ep_{\al}\ra\}$ of the Hamiltonian $H.$
The density matrix described the so-called diagonal ensemble (DE), in which, the 
expectation values of the local operators are given by classical averaging (i.e., with real positive weights -- no interference) over their expectation values on the eigenstates of $H.$ Thus, in some sense, it is sufficient to know the properties of the eigenstates of $H$ and their initial weights to determine the long-term behaviour.

The question then is, whether dropping the phases and time dependence is a valid approximation at late times (clearly it cannot be so at early times where the state is still closed to the initial state which can have a strong phase coherence in the eigen-basis $\{|\ep_{\al}\ra\}$). Loosely speaking, at late times, as $t \to \infty,$ the phase $t(\ep_{\al^{'}} - \ep_{\al})$ oscillates increasing faster, and unless $\al =\al^{'},$ the contribution is cancelled out 
when summed over $\al,\al^{'}$ as they pick up opposite signs randomly -- just in the spirit 
of Riemann–Lebesgue lemma for oscillatory integrals~\cite{Grdstn_Rhyzg}. This process of losing phase information is often referred to as dephasing. 

It should be noted that this dephasing requires significant contributions in the sum in Eq.~\ref{O_t} for several close by values of $\al$ and $\al^{'}$. If there are only a few terms on the right hand side of Eq.~\ref{O_t}, then one obtains a sustained oscillation with 
a few frequencies, and not a steady state described by a DE. This happens if either the
initial state expanded in terms of the eigen-basis $\{|\ep_{\al}\ra\}$ has only a few terms, or $\Ob$ is a non-local projector, which, when applied on $|\psi(0)\ra$ projects out only a few eigenstates. With a typical initial state and a local operator, none of the above is likely 
to happen for any physical (local, short-ranged) Hamiltonian, and hence one can expect a 
diagonal ensemble at late times. The locality of $\Ob$ is hence a {\it sufficient condition} for
getting DE for a generic initial state at late times.

A more elaborate analysis shows, that for a generic $H$ and a local observable $\Ob,$ the state of the system at late times is extremely well approximated (indistinguishable within any realistic experimental resolution) by the DE for overwhelming majority of time during its evolution~\cite{Riemann}. \\

This is already a heavily approximated description, since we have thrown away the information in the phases and the cross terms, however, this still requires knowledge of exponentially many (in $N$) initial information (the weights $|C_{\al}|^2$s) to construct the ensemble. Now if additionally assume that the expectation
value of any local observable $\Ob$ over the energy eigenstates is a 
smooth function of the energy, then for an initial state with
a very narrow energy variance, it is easy to see that the diagonal ensemble is nothing but the microcanonical ensemble for the whole system~\cite{Rigol_Nature}:
\beq 
\Ob_{_{DE}} = \sum_{\al}|C_{\al}|^{2}\la \ep_{\al}|\Ob|\ep_{\al}\ra
\approx \la \ep_{\bar{\al}}|\Ob|\ep_{\bar{\al}}\ra (\bar{E}),
\label{O_ETH}
\eeq
\noi
where $\bar{E}$ is the mean energy of the initial state. This is obtained
by assuming the width of the initial state is very narrow in energy (and
hence, by ETH, in every local observable), and we can approximate all
the expectation values within the sum by that for a single eigenstate $|\ep_{\bar{\al}}\ra$ with energy $\approx \bar{E}.$ This implies,
the effective steady density operator $\rho_{sub}$ for a finite 
sub-system at late times will be given by
\beq
\rho_{sub}(t\to\infty) \approx Tr_{_{Env}}\left[|\ep_{\bar{\al}}(\bar{E})\ra\la \ep_{\bar{\al}}(\bar{E})|\right],
\label{DE_rho_MicroCan}
\eeq 
\noindent
where the operation $Tr_{_{Env}}{[~]}$ implies tracing over the rest of the 
system leaving out the finite sub-system. The ``$\approx$" symbol can be
replaced by the ``$=$" sign in the thermodynamic limit.
Now assuming that energy is the only quantity without fluctuations for the whole system when it is in the eigenstate $|\ep_{\bar{\al}}(\bar{E})\ra$, 
the only relevant independent constraint for the sub-system distribution $\rho_{sub}$ is that the energy  
of the sub-system value should
be fixed to sharp value $\bar{E}_{sub}$ consistent with $\bar{E}.$ Then the maximization of Von Newman entropy immediately 
gives (see, e.g.~\cite{Jaynes})
\beq
\rho_{sub} (t\to\infty) =   e^{-\beta~H_{sub}}/Z_{sub},
\label{DE_rho_Canon}
\eeq 
where $H_{sub}$ is the bulk Hamiltonian of the sub-system, and $Z_{sub}$
is its partition function. The effective temperature can be determined
self-consistently from the equation
\beq
Tr[\rho_{sub}~H_{sub}] = \bar{E}_{sub}.
\label{self_consistent}
\eeq 
It is important to note that obtaining the canonical ensemble from the microcanincal one rests on the assumption that the interactions are short-ranged, and while estimating energy we can neglect the boundary effects in surface to volume ratio, which vanishes in the thermodynamic limit. Hence this does not apply
prima facie to systems with sufficiently long-ranged interactions. \\

For an integrable system with $L \sim O(N)$ LCQs, the late time behaviour will hence be given by a GGE:
\beq
\rho_{sub}^{^{GGE}} (t\to\infty) =   e^{-\sum_{k}\lambda_{k}~{\cal I}^{k}_{sub}}/Z_{sub},
\label{DE_rho_Canon_GGE}
\eeq 
\noi
where ${\cal I}^{k}_{sub}$ are LCQs contained within
the sub-system.
\\

Next we discuss the basics of Floquet quantum matter and how to apply the above to characterize its late time behaviour.

\section{Floquet Quantum Matter and its Statistical Mechanics}
\label{Flq_Stat_Mech}

\subsection{Steady States of Floquet Matter}
We begin with the precise definition of Floquet quantum matter.
It is essentially a closed quantum many-body system driven periodically in time. 
In most cases, it is considered to be in a pure state at all time.
The dynamics of the entire system is also deterministic. The statistical description hence applies, as usual, only to a finite sub-system of the whole system, or, in other words, only for describing local operators as it is for the equilibrium 
(time-independent $H$) case discussed above. \\

Given the above definition, closed Floquet matter is inherently out-of-equilibrium, hence applying
the idea of equilibrium statistical mechanics to it might apparently look antithetical.
So it did till the beginning of the last decade -- to the least,
one has to have a non-trivial stable steady state to describe, but the common wisdom was, a periodically driven system with many degrees of freedom should keep on absorbing energy without bound (as, for example, suggested by Fermi's Golden rule) and will tend to a completely featureless random state asymptotically, since energy itself is not conserved.

However, it turns out not to be the entire story -- it was shown, the polarization of an infinite integrable spin chain can remain dynamically frozen arbitrarily close to its initial state {\it at all time} under simple periodic drive, if the drive is sufficiently strong and fast~\cite{AD-DMF}. The freezing is a non-monotonic function of the drive frequency and the amplitude, and is almost perfect under certain drive conditions. We will see the most interesting aspect of the freezing is probably that it is approximate, but perpetually stable. But from statistical mechanics point of view, the most important aspect is, the freezing is {\it independent of the initial state}. The phenomenon thus heralded the existence of a 
equilibrium ensemble-like description of Floquet quantum matter. \\

\noi 
The phenomenon was later dubbed as dynamical many-body freezing (DMF) ~\cite{SB_AD_SDG_2012,Mahesh_DMF_2014,Kris-Periodic,Russomanno,AD_RM_Switching,Bastidas_Ising,Bastidas_Dicke,Luca_Polku,Diptiman_Adhip_DL_1,Bhaskar_Scar,Roopayan_DMF},and is basically a many-body version
of long-known single-particle phenomenon of dynamical localization~\cite{Dunlap_Kenkre,CDT_PRL}. Most importantly, DMF has recently been shown to persist for a large class of non-integrable interacting systems with the undriven part containing Ising or two-body Heisenberg interactions of virtually any kind, and on lattices of any dimension~\cite{Asmi_PRX}. Stability of DMF entails an approximate emergent conservation law. Subsequent works showed, non-trivial steady states are generic to non-interacting Floquet matter due to presence of exact conservation laws~\cite{PGE}.
\\

\subsection{From Floquet to Quantum Quenches}
The connection between the late-time behaviour of quantum Floquet matter and quantum quenches with time-independent Hamiltonian (and hence with the ensemble picture describing their late time behaviour) becomes apparent if one observes a Floquet system stroboscopically. Floquet quantum matter is described by a time-periodic many-body Hamiltonian
\beq
H(t+T) = H(t).
\label{H_T}
\eeq 
\noindent The time-evolution operator can be cast in the form 
(Floquet-Bloch theorem, see, e.g.~\cite{Mostafazadeh})
\bea  
U(t; 0) &=& \Omega(t)~e^{-iH_{eff}t}, ~ {\rm with} \non \\ 
\Omega(t+T) &=& \Omega(t), ~ \Omega(0) = \Omega(\ell~T) = {\mathcal I}
\label{U_T}
\eea  
\noi where $U(T; \epsilon)$ denotes the time evolution operator for evolution from $t=\epsilon$ to $t=\epsilon +T.$
$H_{eff}$ is a time-independent hermitian operator (often called 
Floquet Hamiltonian), ${\mathcal I}$ the Identity operator, and $\ell$ an integer. 
All the information of the stroboscopic dynamics is stored in $H_{eff},$ and $\Omega(t)$ contains the information
about the dynamics within a period (so-called micromotion).
Now if one focuses
only on stroboscopic observations at discrete moments given by $t = \ell~T$
(i.e., after $\ell$ complete cycles), then the states will be given by
\beq 
|\psi(\ell~T)\ra = [U(T; 0)]^{\ell}|\psi(0)\ra = e^{-iH_{eff}\ell~T}|\psi(0)\ra.
\label{psi_nT}
\eeq 
\noi
This would also exactly be the state at $t=\ell~T$ 
if the evolution was carried out by the time-independent 
Hamiltonian $H_{eff}.$ Eq.~\ref{psi_nT} thus maps the evolution of a Floquet system to a quench problem with the time-independent Hamiltonian $H_{eff}$ under stroboscopic observations,
where 
\beq 
H_{eff} =\frac{i}{T} \ln{[U(T; 0)]}.
 \label{Flq_M}
\eeq  
$H_{eff}$ being hermitian, its eigenstates $|\mu_{\al}\ra$ (Floquet states) with eigenvalues $\mu_{\al}$ can be ortho-normalized to form a complete basis known as Floquet basis. Note that $|\mu_{\al}\ra$  are also eigenstates of $U(T; 0)$ with eigenvalues $e^{-i\mu_{\al} T}.$
These basis plays a role similar to that played by the energy eigen-basis in time-evolution under a time-independent Hamiltonian. \\

We are now equipped to apply all the constructs and concepts discussed in the previous section (Sec.~\ref{Sec:Eqlb_Stat_Mech}) to describe the late time behaviour of Floquet matters. The rest of the review is tailored to answer questions laid down in the beginning of this subsection.
\subsection{Periodic Steady State and Synchronization}
The first question concerns emergence of a steady state
of some sort in Floquet matter at late times. The argument of dephasing in Sec.~\ref{Sec:Eqlb_Stat_Mech} implies that at late times, the
 a generic Floquet system starting from a generic initial state 
will execute a periodic variation with time, synchronized with the drive. 
To see this, let us consider a $T-$periodic drive as discussed above. 
Let us 
expand a generic initial state
\beq 
\label{Flq_Expns}
|\psi(0)\ra = \sum_{\al} f_{\al}|\mu_{\al}\ra,
\eeq 
\noi
where $|\mu_{\al}\ra$ are the Floquet eigenstates. Then, since when observed at instants $t = \ell~T$ the
state will be exactly the one obtained due to evolution with the time-independent Hamiltonian $H_{eff}$ corresponding to the $U(T; 0)$ for the given problem,
(see Eqs.~\ref{psi_nT}-\ref{Flq_M}) we expect, following Eq,~\ref{DE_O}, 
for $t = \ell~T$
\beq 
\la\Ob(\ell\to\infty)\ra \approx \Ob_{_{DE}}^{(0)} = \sum_{\al}|f_{\al}|^{2}\la \mu_{\al}|\Ob|\mu_{\al}\ra.
\label{Flq_DE_O}
\eeq
\noi
Here the superscript``$(0)$" indicates that we have considered the stroboscopic observation series $t = [~0, T, 2T, ...\ell~T ...~]$. For this series, as we see, $\Ob^{(0)}(\ell~T)$ is expected to attain a time-independent value. Now instead of the series $t = \ell~T,$ we could have taken the series $t = \epsilon + \ell~T,$ where $0 < \epsilon < T,$ and carry out the above arguments exactly in the same way, only, now taking $t = \epsilon$ as our starting instant -- $|\psi(0)\ra$ is replaced by $|\psi(\epsilon)\ra,$ and $H_{eff}$ by 
$H_{eff}({\epsilon}) = \frac{i}{T}\ln{[U(T;\epsilon)]}.$ This immediately implies at late time, for $t = \epsilon + \ell~T$
\beq
\la\Ob(\ell\to\infty)\ra \approx \Ob_{_{DE}}^{(\epsilon)} = \sum_{\al}|f_{\al}|^{2}\la \mu_{\al}(\epsilon)|\Ob|\mu_{\al}(\epsilon)\ra,
\label{Sync}
\eeq 
\noi for every $0 <\epsilon < T.$ 
Here $|\mu_{\al}(\epsilon)\ra$ are the eigenstates of 
$U(T; \epsilon)$ with eigenvalues $ \mu_{\al}$ (the eigenvalues are independent of $\epsilon$).
Eq.~\ref{Sync} implies that the variation
$\Ob_{_{DE}}^{(\epsilon)}$ 
of the DE average of a local observable $\Ob$ 
over a period given by $\delta + \epsilon$
as $\epsilon$ runs from $0$ to $T,$ will be the same
over the next period $\delta + T + \epsilon$ 
as $\epsilon$ runs from $0$ to $T.$ In other words, $\la\Ob\ra(t)$ get synchronized with the drive 
at late times -- this is the Floquet ``steady state." 

\subsection{Floquet Dynamics {\it vs} Quantum Quench with a Local Hamiltonian: Looking Through the Magnus glass} 
From the viewpoint of a late-time behaviour, the main difference between quench
with a natural physical (local, short-ranged) time-independent Hamiltonian $H_{stat}$
and stroboscopic Floquet dynamics with an $H_{eff}$ lies in some crucial differences 
possible between the structure of $H_{eff}$ and $H_{stat}$. Even when the periodically driven
Hamiltonian $H(t)$ (which gives rise to $H_{eff}$) is a 2-local, short-ranged Hamiltonian,
$H_{eff}$ can have the following exotic properties. (a) $H_{eff}$ can be long-ranged, even infinite ranged.
(b) For low drive frequency $H_{eff}$ can contain substantially strong multi-body
    ($k-$local, with $k > 2$) interactions, and can even be non-local. 
These properties can naturally make the eigenstates $|\mu_{\al}\rangle$ of $H_{eff}$
very different from the eigenstates of a standard $H_{stat},$ and hence can produce a diagonal ensemble with drastically different properties from a diagonal ensemble with $H_{stat}.$ \\
\begin{figure*}[htb]
  \begin{center}
    \includegraphics[width=0.9\linewidth]{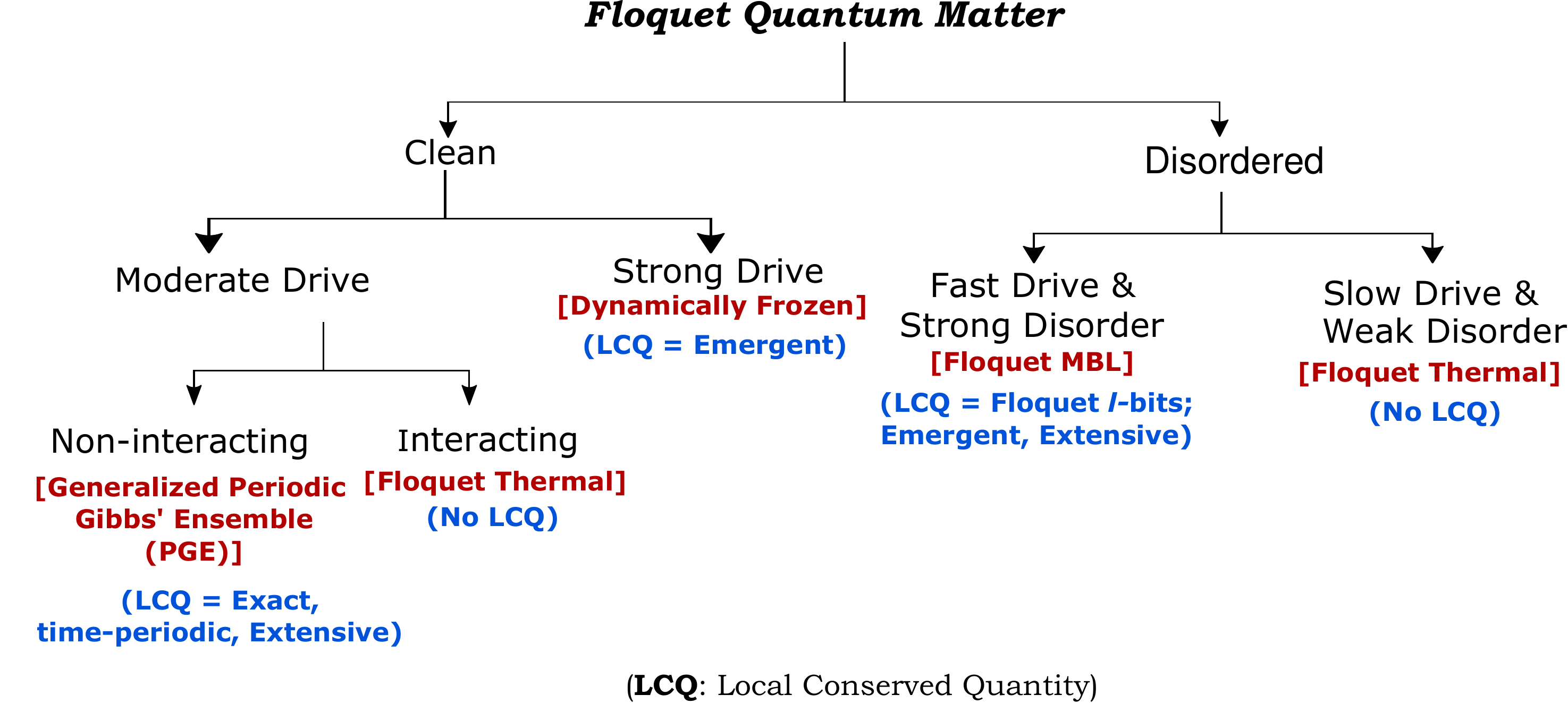}
  \end{center}
  \caption{A Broad Classification of Floquet Quantum Matter indicating their stability and Floquet conservation laws. 
  Here ``Floquet thermal" means locally infinite-temperature like state, and "extensive" means there are as many LCQs as
  the number of degrees of freedom.
  The classification is not complete. There can be stable 
  Floquet phases in non-interacting disordered Floquet systems,
  but the nature of the conserved quantities are yet to be studied.
  }
  \label{Classification}
\end{figure*}
There are several ways of calculating $\heff$ (see, e.g., \cite{Goldman_Dalibard,Anatoli_Rev,Andre_Anisimovas_Rev}).
A rather transparent and popular one is the Magnus expansion~\cite{Magnus} (see also \cite{Anatoli_Rev}). This is a series expansion for $H_{eff}$ in $1/\omega$ ($\omega = 2\pi/T$ is the drive frequency). 
Magnus expansion is given in terms of nested commutators between the drive Hamiltonian $H(t)$ at different instants as follows. 
\begin{eqnarray}
	H_{eff} &=& \sum_{n=0}^{\infty} H^{(n)}_{eff}, ~ {\rm where} \nonumber \\
	H^{(0)} &=& \frac{1}{T}\int_{0}^{T} H(t) dt \nonumber \\
	H^{(1)} &=& \frac{1}{2\!T(i\hbar)}\int_{\epsilon}^{\epsilon+T}dt_{1} 
	\int_{\epsilon}^{t_{1}}dt_{2}[H(t_{1}),H(t_{2})]\nonumber \\
	H^{(2)} &=& \frac{1}{3\!T(i\hbar)^{2}}\int_{\epsilon}^{\epsilon+T}dt_{1} 
	\int_{\epsilon}^{t_{1}}dt_{3}\left([H(t_{1}),[H(t_{2}),H(t_{3})]]\right. \nonumber \\
	&+& \left.[H(t_{3}),[H(t_{2}),H(t_{1})]]\right) \dots 
\label{Def:Magnus}
\end{eqnarray}
\noi An interesting thing about Magnus expansion is, it is hermitian at every order. Magnus expansion
provides a way to derive a sufficient condition for absence of unbounded heating and stability of Floquet matter. \\

If $H(t)$ is a many-body Hamiltonian, then in the thermodynamic limit the radius of
convergence of the above expansion is usually zero~\cite{Kato_Book}. However, under special circumstances
this series can be shown to converge or to be asymptotic. Those are the cases where the stability of a 
Floquet phase can be guaranteed. The conserved quantities relevant for
constructing the statistical mechanics for a Floquet system  are local operators which
commutes with $H_{eff}.$ Also, the symmetries of $H_{eff}$ are the symmetries that 
characterize the Floquet phases. \\

\noi
In the rest of the article we will review the existence and nature of Floquet ensembles and the conservation laws characterizing them. Based on the nature of stability, three broad classes of  Floquet matter have been discussed most widely -- the non-interacting, the clean interacting, and the MBL, to which we will limit our discussions.

At this point it seems useful to clarify two important concepts, namely,
ergodicity and emergent conservation laws, especially the sense 
in which they are used here. \\

First, we are using the word ergodicity in a more inclusive sense here
-- by it we mean the dynamics and/or eigenstate properties (e.g. in ETH) are such that
the correct statistical distribution (for the sub-system observables) 
maximizes the entropy respecting all exact conservation laws for local operators. \\

\noi
Second is the idea of emergent conservation laws. In the Floquet context, exact conservation laws may trivially be inherited from the
drive Hamiltonian $H(t)$ broadly through three different ways -- either due to the symmetries in $H(t)$, 
or owing to its integrability, or due to disorders (e.g., $\ell-$bits in MBL systems, see e.g.~\cite{Dima_RMP,Imbrie_Review}). In addition to those, there may arise further constraints due to the dynamics, 
which can sometimes take the form of approximate LCQs. These emergent LCQs have no trivial
origin that can be guessed looking at $H(t)$ at any given instant. They can often leave legible foot-prints on
$H_{eff},$ but since convergent series for $H_{eff}$ usually do not exist for any non-trivial many-body system,
they are hard to guess even from $H_{eff}$ in general. Hence, while constructing the ensemble by maximizing the entropy, those are left out (only exact conserved quantities are taken into account). 
Hence they break the ergodicity of a system in a surprising manner. An algorithm to
identify the emergent conservation laws and incorporating them in constructing statistical distributions is an important open problem in the way of constructing a complete statistical mechanics for describing
late time behaviour of complex dynamical systems. Until this problem is
solved, we are in an rather uncomfortable position: our ensemble
description could be incomplete, and ergodicity could be found broken
in an unexpected way by some ad hoc emergent conservation law. We will see, these emergent conservation laws are responsible for non-trivial late-time behaviour of interacting Floquet matter initialized to generic initial conditions. Fig.~\ref{Classification} charts out a broad (though incomplete) classification of
Floquet matters based on their stability and associated LCQs. We will roughly follow the flow of the diagram for the rest of the review.
\\

\section{Non-interacting Floquet Matter}
\label{Sec:Floquet_PGE}


\begin{figure*}[ht]
  \begin{center}
    \includegraphics[width=0.9\linewidth]{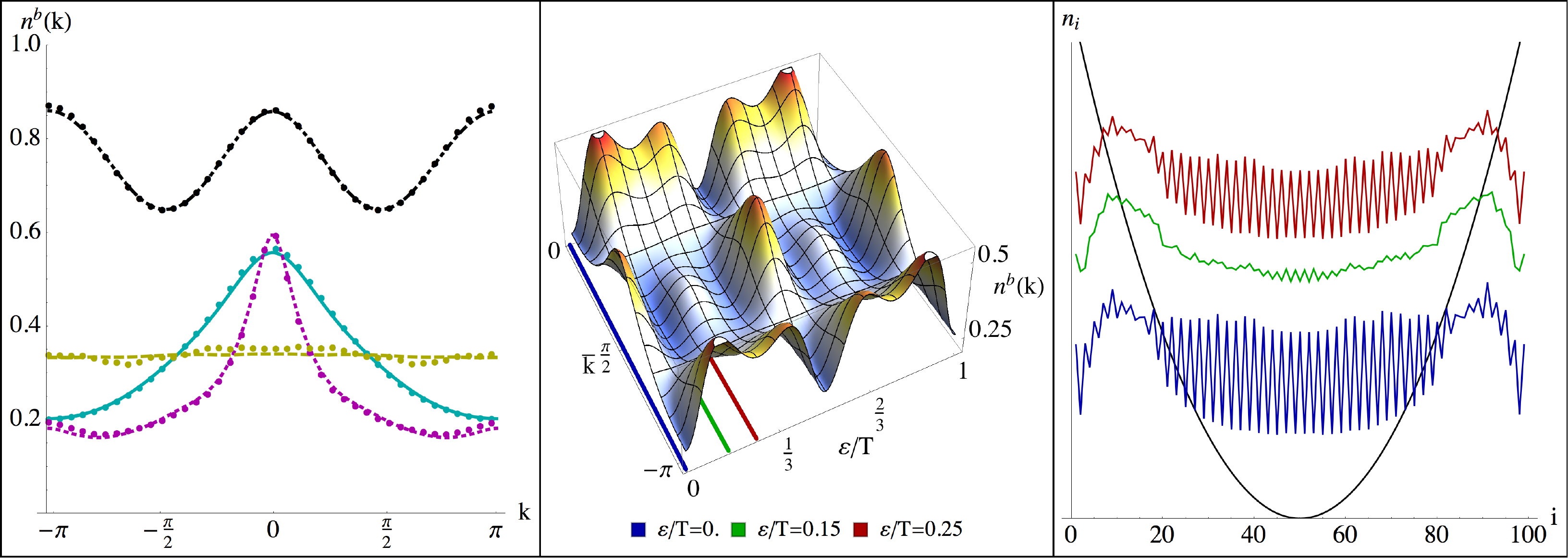}
  \end{center}
  \caption{
  Characterisation of the synchronised steady state for Hamiltonian $H_{b}(t)$ (Eq~\ref{eq:hcb}) for $N=100$.
        {\bf Left frame}: Stroboscopic momentum distribution, 
        $\hat{n}(k)=L^{-1} \sum_{i,j}\hb^\dagger_i\hb_j
          \exp(-2\pi i k(i-j)L^{-1})$. The points correspond to snapshots
		of the dynamical evolution (exact numerical results) at late times ($t=490T$) for $L=200$,
		while the continuous lines correspond to the PGE prediction.
		The parameters
        $(\Delta,\delta J,\omega,\nu)$ are $(0.6,0.5,1.6,3/4)$ (black,
        dot-dashed), $(4,0.5,1.5,1/3)$ (yellow, dashed), $(4,0.75,2,1/3)$ (cyan,
        full), $(0.6,0.5,2,1/4)$ (magenta, dotted) and $\eps=0$. 
		The next two panels correspond to 
        the parameters for the cyan full line. 
		{\bf Middle frame}: 
		Expectation value of the momentum distribution $\hat n(k)$ within a
        single period in the synchronized state as a function of the time 
        $\epsilon$ within the period. 
        The three lines on the time-momentum plane indicate the 
        times $\epsilon/T=0, 
        0.15, 0.25$ for which density distributions $\hat{n}^b_i=\hb^\dagger_i\hb_i$ 
        are shown in the {\bf Right frame} (Taken from ~\cite{PGE}). 
  \label{fig:final-Values-And-Period}
}
\end{figure*}

\subsection{The Generalized Periodic Gibbs' Ensemble (PGE) and its conservation laws}
The stability of the polarization of an integrable Ising chain~\cite{AD-DMF} (which can be mapped to free fermions) under periodic drive raised the question of the role of integrability and the possibility of existence of some sort of conservation laws in such systems, that prevent unbounded heating (though energy itself is not conserved). It was shown~\cite{PGE} that a general integrable system constituting of free fermions (or one that can be reduced to free fermions), when driven periodically, indeed hosts an extensive number of $T$-periodic LCQs $\Ik(t+T) = \Ik(t).$ A
general non-interacting fermionic Hamiltonian of the form
\begin{equation}
  \hat{H}(t) = \sum_{i,j}\left[
      \ha_i^\dagger \cM_{i,j}(t)\ha_j + 
      \ha_i^\dagger \cN_{i,j}(t) \ha_j^\dagger + \mathrm{h.c.}
    \right],
  \label{eq:quadratic}
\end{equation}
\noi
was considered,
where $\ha$ are fermionic, and $\cN, \cM$ are complex matrices.
For one dimensional lattice, $\ha$ can also be hardcore bosons.
Now the key observation is, given the nested commutator structure of the
Magnus expansion of $H_{eff}$ (Eq.~\ref{Def:Magnus}), if $H(t)$ is bi-linear
in fermions, so if $H_{eff}.$ 
Thus, in principle, using a unitary transformation
(see, e.g.,~\cite{Sei_Book}) one can always reduce $\heff$ 
to a free fermionic Hamiltonian of the form
\begin{equation}
  \heff=\sum_{k=1}^N \omega_k \tc^\dagger_k\tc_k
  \label{eq:heff}
\end{equation}
\noi
Note that the operators $\tc$'s can be quite complicated in terms of the
original degrees of freedom, but the final result is still bi-linear 
since the algebra closes in each order of the expansion.
The operators 
\beq
\Ik(0) = \tc^\dagger_k\tc_k, ~ k=1,N
\label{GGE_Ik}
\eeq 
are relevant (2-local) {\it periodically conserved} 
quantities for the stroboscopic steady series $t=\ell~T.$  
One can explicitly construct time-periodic quantities, that will be stroboscopically
conserved, i.e. $\Ik(t+T) = \Ik(t),$ and hence will be useful for constructing the
synchronized ensemble. The prescription is
\bea  
\la \psi(t)|\Ik(t)|\psi(t)\ra &=& \la\psi(0)|\Ik(0)|\psi(0)\ra 
\label{GGE_Ik_strobo}
\eea 
\noi
for all $t.$ This completes the list of ingredients to construct the ensemble that 
represents the late-time synchronized state of the integrable Floquet matter described above. 
We then just follow the statistical mechanics
prescription given in Eq.~\ref{DE_rho_Canon_GGE} to
construct the ensemble.
The ensemble thus constructed is termed as (generalized) Periodic Gibbs' Ensemble (PGE)~\cite{PGE}. It is given by the
density operator
\begin{equation}
  \label{eq:pge-rho}
  \hat{\rho}_{_{PGE}}(t)=Z^{-1}\exp\left(-\sum_{k}\lambda_{k}\Ik(t)\right)
\end{equation}
\noindent
where the $\lambda_k$s are fixed by requiring that
\bea 
\bra{\psi(0)}\Ik(0)\ket{\psi(0)}&=& {\rm Tr}\left[ \hat{\rho}_{_{PGE}(0)} 
\Ik(0)\right] ~ {\rm and} \non \\
Z &=& \left(\mathrm{tr}\,\hat\rho_{_{PGE}}(t)\right)^{-1},
\label{Z_PGE}
\eea 
\noi
where $Z$ is a time-independent normalization factor analogous to the
partition-function. \\

The density operator $\hat{\rho}_{_{PGE}}(t)$ has the following two properties.
\noi
(a) It correctly gives the conserved quantities.
\bea 
{\rm Tr}\left[\hc^\dagger_k\hc_k^{'}\hat{\rho}_{_{PGE}}(t)\right]
=\delta_{k,k^{'}}\bra{\psi(t)}\Ik(t)\ket{\psi(t)}. \non
\label{PGE_Avg_Chck}
\eea   
\noi  
(b) Since the $\Ik(t)$ are periodic in time, $\hat{\rho}_{_{PGE}}$ is itself manifestly periodic in time:
$\hat\rho_{_{PGE}}(t)=\hat\rho_{_{PGE}}(t+T),$ in conformation the synchronized steady state.
The description hence requires the values of only $N$ conserved quantities, 
rather than $~e^N$ numbers to specify the initial conditions exactly.
\\

We end this section by demonstrating the predictive power of PGE.
The accuracy of the PGE is demonstrated by comparison with exact numerical result
for an experimentally relevant system of hardcore-bosons on a one-dimensional lattice
given by the Hamiltonian~\cite{PGE}
\beq 
 \hat{H}_{b}(t)=-\frac{1}{2}\sum_{i=1}^{N} J_i(t) \hbd_i \hb_{i+1}+\mathrm{h.c.}
    +\sum_i V_i(t)\hbd_i \hb_i.
    \label{eq:hcb}
\eeq 
Here $\hb_i$ are hardcore-bosons (HCBs) obeying bosonic commutation relations off-site, $\left[\hb_i,\hb_j^\dagger\right]=\delta_{i,j}$ with
additional hardcore condition $\hb^2_i=0$ onsite.
A Jordan-Wigner transformation, $\hb_i=\ha_i \prod_{j<i}(-1)^{\hat{n}_j}$ with 
$\hat{n}_j=\hb_j^\dagger \hb_j=\ha_j^\dagger\ha_j$, maps
$\hat{H}_{b}(t)$ to Eq.~\eqref{eq:quadratic} with
$\cM_{i,j}(t)=-\frac{1}{2}J_i(t)\left(\delta_{i+1,j}+\delta_{i-1,j}\right)+\delta_{i,j}V_i(t)
$, $\cN_{i,j}=0$ and fermionic commutation relations for the $\ha$. \\

The driven terms are, a time-dependent super-lattice potential superposed on a
quadratic potential,
$V_i(t)=\frac{1}{2}\left(\left(i-L/2\right)/\ell_{ho}\right)^2 +\Delta
(-1)^i\cos\left(\omega t\right)$ and a time-dependent hopping amplitude $J_i(t)=J+\delta
J\cos(\omega t)$ with $\omega=2\pi/T$.
The protocol used here consists of preparing the system in the ground state 
in the presence of a harmonic potential
$V^{(0)}_i=\frac{1}{2}\left(\left(i-L/2\right)/\ell_{ho}\right)^2$, fixing
$\ell_{ho}=N$. This allows taking the thermodynamic limit, 
since for large total number of particles the dimensionless 
parameter~\cite{RigolUniversal} $\tilde{\rho}=N_b/\ell_{ho}$ plays a role analogous to the density in 
the uniform limit. Results with different system sizes but constant $\tilde{\rho}$ are therefore
comparable.

At time $t=0$, the driving is switched on so that the total Hamiltonian is
$\hat{H}_{b}(t)=
  -\frac{1}{2}J\sum_i \hbd_i \hb_{i+1}+\mathrm{hc}+\sum_i V_i(t)\hbd_i \hb_i$
with $V_i(t)=V^{(0)}_i+\Delta (-1)^i\cos\left(2\pi t/T\right)$.
Considering the experimentally accessible momentum distribution
of the bosons, $\hat{n}^{(b)}(k)=L^{-1} \sum_{i,j}\hb^\dagger_i\hb_j\exp(-2\pi
k(i-j)L^{-1})$ a numerical method consisting of solving the
fermionic time-dependent problem and, in the end, inverting the Jordan-Wigner 
transformation was employed. \\

We demonstrate various periodic states
corresponding to different sets of values of the
parameters of the model. The leftmost panel of Figure~\ref{fig:final-Values-And-Period}
shows snapshots of the PGE momentum distribution 
\beq 
{\rm Tr}\left[\hat{\rho}_{_{PGE}}\hat{n}^{(b)}(k)\right]
\label{PGE_Momntm_Dstrb}
\eeq
\noi at late time (after $490$ cycle). The dashes lines are the PGE prediction, while the points are exact numerical
results. The agreement between the both resoundingly confirms the validity of the PGE picture at late times. \\

It is to be noted that at low-frequency regime, as expected, the corresponding time-averaged Hamiltonian~\cite{Andre_Anisimovas_Rev} is not an appropriate description. As a striking illustration, the black line shows a momentum distribution with peaks at the edges of the Brillouin zone. The central panel shows the time evolution of the momentum distribution over an entire  period corresponding to the parameters associated with the cyan line in the left frame. Of course, the system evolves through states in which the momentum is peaked 
at different locations of the Brillouin zone. 
Finally, the rightmost panel shows three snapshots of the
density distribution of the same system at times indicated by the coloured 
lines (black, green and red) in the central panel. The high frequency spatial oscillations and the peaking of the density at the edges is also very different from what would be obtained had the system been well-described by a
time-averaged Hamiltonian, since the time-averaged potential (shown in black) is smooth and peaks at the edges.
\\

\subsection{Dynamical Many-body Freezing (DMF) and An Emergent Conservation Law in Non-interacting Floquet Matter}
\label{DMF_Integrable}
\begin{figure}[h!]
  \begin{center}
    \includegraphics[width=0.85\linewidth]{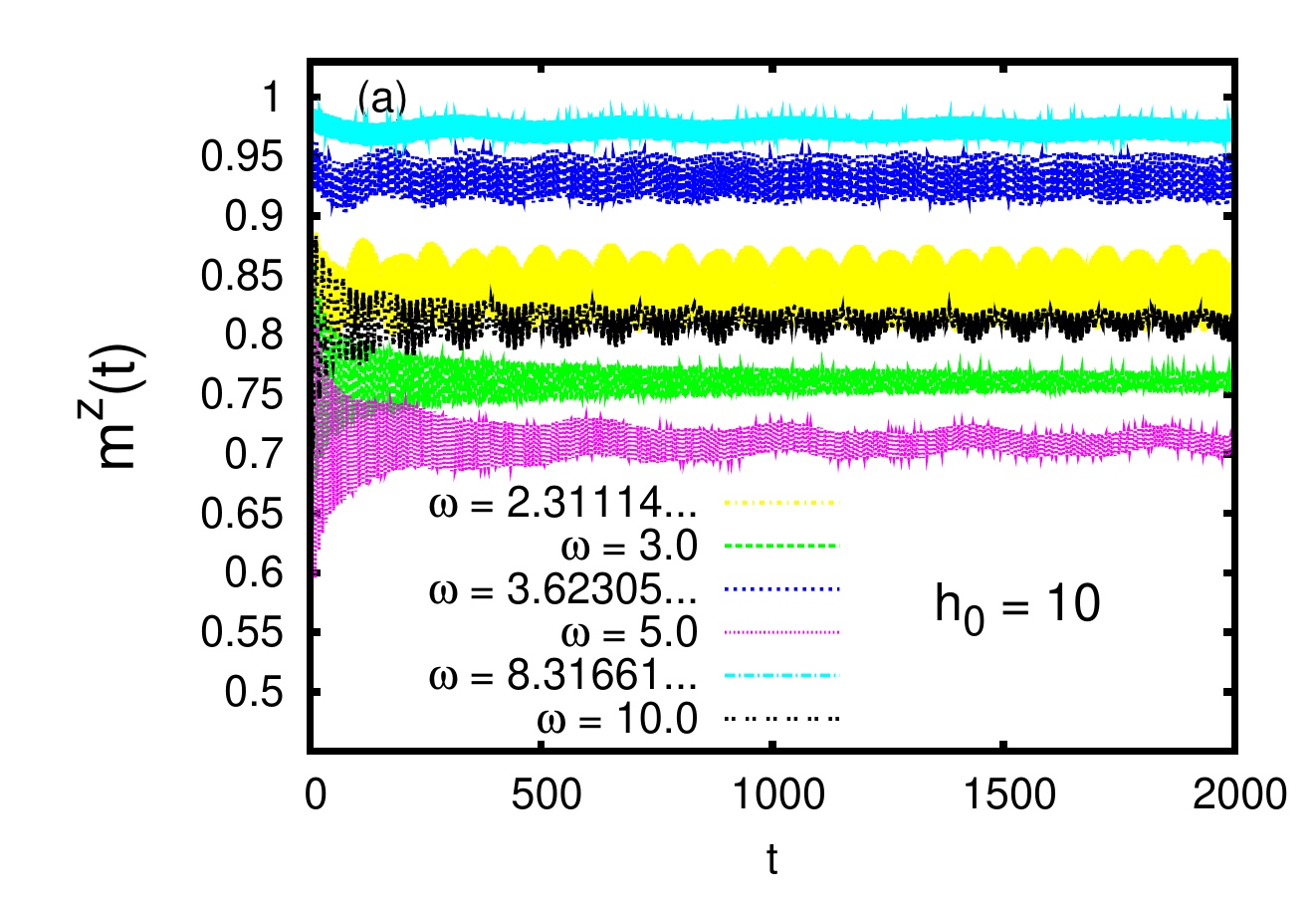}
  \end{center}
  \caption{
  {\bf Breaking of ergodicity by an {\it emergent} conservation law}: $m^z$ is not among the exact conserved
  quantities that participates in the PGE for the system, yet 
  it enjoys a perpetual approximate conservation, strongly constraining the dynamics on top of
  the exact conservation laws participating in the PGE.
  Time evolution of the approximately conserved $z-$polarization $m^z$ for several drive cycles ($N=100$). The steady time-average of $m^z$ persists undiminished even after infinite drive cycles. The freezing was checked numerically to persist undiminished 
  up to $L=10^5$ for up to $\ell = 10^4$ cycles (unpublished), from which is consistent with the 
  analytical prediction for the freezing to persist in
  the $N\to\infty$ limit and after infinitely many cycles. (Taken from~\cite{AD-DMF}).
            }
  \label{DMF_Time_Evol}
\end{figure}

For non-interacting Floquet matter, as we have just seen, there are $N$ number of relevant $T-$periodic conserved quantities. Hence ergodicity in such system would mean a distribution that respects all those conservation laws, and no other constraints -- exact or approximate -- at late times. This gives us the PGE (Eq.~\ref{eq:pge-rho}). \\

However, already before the inception of the ideas of PGE and its exact conservation laws, it was shown, the ergodicity thus defined can be broken by {\it emergence} of a conserved quantity when the drive is sufficiently strong~\cite{AD-DMF}. The following strongly driven integrable quantum Ising chain was 
considered 
\begin{equation}
H(t) = -\frac{1}{2}\left[J\sum_{i=1}^{N} \sigma_{i}^{x}\sigma_{i+1}^{x}
+ h_{z}(t)\sum_{i=1}^{N} \sigma_{i}^{z}\right],
\label{H-Ising}
\end{equation}
\noindent
($h_{z}(t) = h_{0}\cos{(\omega t)}$ 
is the driving field and
$\sigma^{x/z}_{i}$
are $x/z$ component of the Pauli spin sitting at site $i$).
The above Hamiltonian can be reduced to that of free fermions and can be cast in the form of Eq.~\ref{eq:quadratic} using Jordan-Wigner 
transformation (see, e.g.,~\cite{Sei_Book}), hence should be completely
described by a PGE with $N$ exact conserved quantities. However, as
discussed below, that turns out not quite to be the case.
\\

\noi 
In the fast drive regime, 
starting from a highly polarized initial state $(m^z = \frac{1}{N}\sum_{i=1}^{N}\sigma_{i}^{z} \sim 1),$ the polarization $m^z$ acquires an steady average value $Q$ depending on the drive amplitude $h_0$ and frequency $\omega$, about which it fluctuates. In the $L\to\infty$
limit, for average over infinite time, the time-averaged magnetization can be captured under a rotating-wave approximation~\cite{AD-DMF}
by
\beq 
\overline{m^{z}(t)} = Q = \frac{1}{1 + J_{0}\left(\frac{2h_{0}}{\omega}\right)},
\label{Q_Analytic}
\eeq 
\noi
where $J_{0}$ is the ordinary Bessel function of order $0.$
From this, we see, for certain specific values of $\omega$ and $h_0,$ satisfying the peak freezing condition:
\beq
J_{0}\left(\frac{2h_{0}}{\omega}\right) = 0,
\label{Freezing_Cond_Cosine_Drv}
\eeq 
\noi 
the system remains maximally frozen with $Q \approx 1.$ 
Similar freezing is also obtained for square-pulse drive 
$h_{z}(t) = \Gamma_{0}\Sgn [\cos{(\om t)}],$ ($\Sgn$ is the Signum function), applied on the same model -- just the condition for the peak freezing becomes
\beq
\Gamma_0 = k\omega, ~ k = 1, 2, 3 ... ,
\label{Frz_Cond_Sqr_Drv}
\eeq 
In these systems, there is clearly an approximate {\it{emergent}} 
conserved quantity, namely, $m^z.$ This is certainly a strong constraint in additional to the $N$ exactly conserved stroboscopic quantities $\Ik$s and evidently {\it independent} of them. This thus breaks the ergodicity further in this integrable system: the PGE, which does not include $m^z$ by construction, is hence not an adequate local description for the system. We will see, in the presence of non-integrable interactions (quantum-chaotic many-body systems) while the exact local conservation
laws disappear, the emergent conservation laws survive and continues to shape the statistical 
mechanics of the steady state, especially, preventing it from attaining the featureless, locally infinite-temperature like state.
\\

\subsection{Infinite-Temperature like Scenario in Integrable Floquet Matter:}
We conclude this section with interesting results underscoring the
non-triviality of DMF in the above integrable systems. These systems
can actually be driven to locally infinite-temperature
ensemble under slow enough periodic drive~\cite{PGE_Infinite_T}!
Despite the existence of $\sim N$ local conservation laws, 
this happens at low enough $\omega$ when the Magnus expansion fails to converge in any local basis
(i.e., basis which is formed by the complete ortho-normal set of eigenstates of some local operator). 
In that case, though the conserved quantities exist, they are unable 
to constrain the statistical 
distribution $\rho_{_{PGE}}$ (Eq.~\ref{eq:pge-rho}), since in that case one
obtains $\lambda_{k} \approx 0$ for all $k$, rendering $\rho_{_{PGE}} \approx {\mathcal I}$ (the identity operator), representing an infinite-temperature state. \\

Finally, it is noteworthy that phase coherence
throughout the dynamics
is crucial for the stability of the extensive number of conservation 
laws in a non-interacting system. It was shown that even in a non-interacting translationally invariant system, which can be mapped to a bunch of decoupled two-level 
systems in the momentum space, can thermalize to infinite temperature (each of the two level systems approach a state with equal probability for the two levels, and hence of maximum entropy) if the St\"{u}ckelberg-like phases are randomized or discarded by hand after each half-cycle, and transition probabilities are used instead of amplitudes to calculate the probabilities after every cycle~\cite{Mahesh_DMF_2014,Amit_Repeated_Interference}. The locally 
infinite-temperature like state is approached exponentially fast with
number of drive cycles. This shows the stability of the periodic conservation laws $\Ik$ depends crucially on the coherence of the 
repeated quantum interference taking place over cycles.

\section{Interacting Floquet Matter under Moderate Drive: 
Floquet ETH and the Late-time Ensemble}
\begin{figure}[h!]
\includegraphics[width=0.80\linewidth]{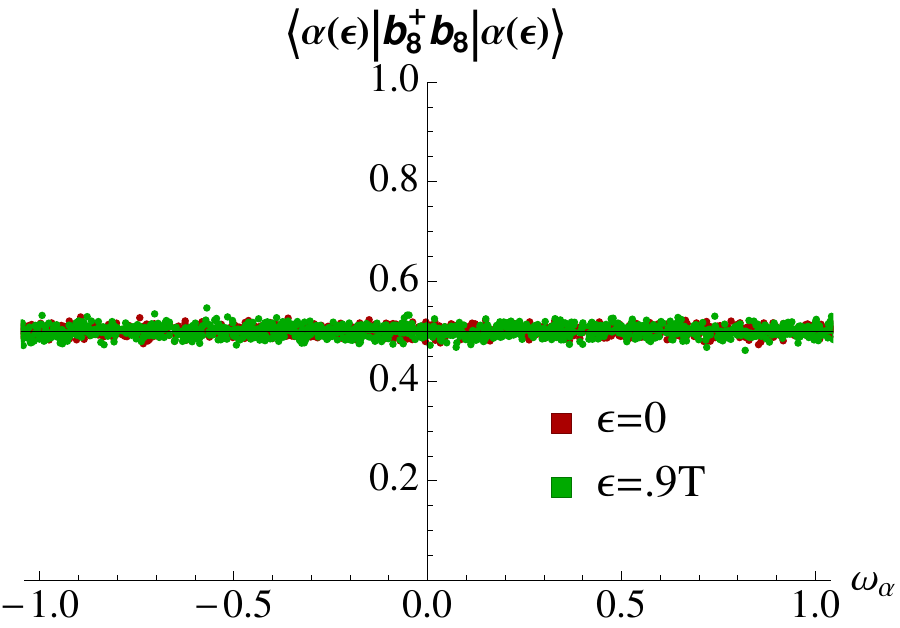}
\caption{An example of the dependence of EEV of an onsite
boson density $b_{8}^{\dagger}b_8,$ on the quasi-energy $\omega_{\al}.$ 
The result corresponds to $u/\omega=1$ and lattice size $L=14$ and particle number $N=7.$
for a Hilbert space dimension $D_{H}=3432$, with parameters
$u=V_1=V_2=J$ and driving frequency $\hbar \omega=h/T=J/4$. Points indicate
expectation value of the density at site $i=8$ in 
an eigenstate $\ket{\alpha(\epsilon)}$ of $H_{eff}(\epsilon)$
versus the state's quasi-energy $\omega_\alpha$ at two different times.
Results for two different stroboscopic series corresponding two $\epsilon = 0$ and $0.9$
are shown. The black line indicates $\mathrm{tr}\left(b^\dagger_8 b_8\right)=N/L=0.5$
(Taken from~\cite{LDM_PRE}).
\label{fig:eev-example}}
\end{figure}
\begin{figure}[h!]
\includegraphics[width=0.7\linewidth]{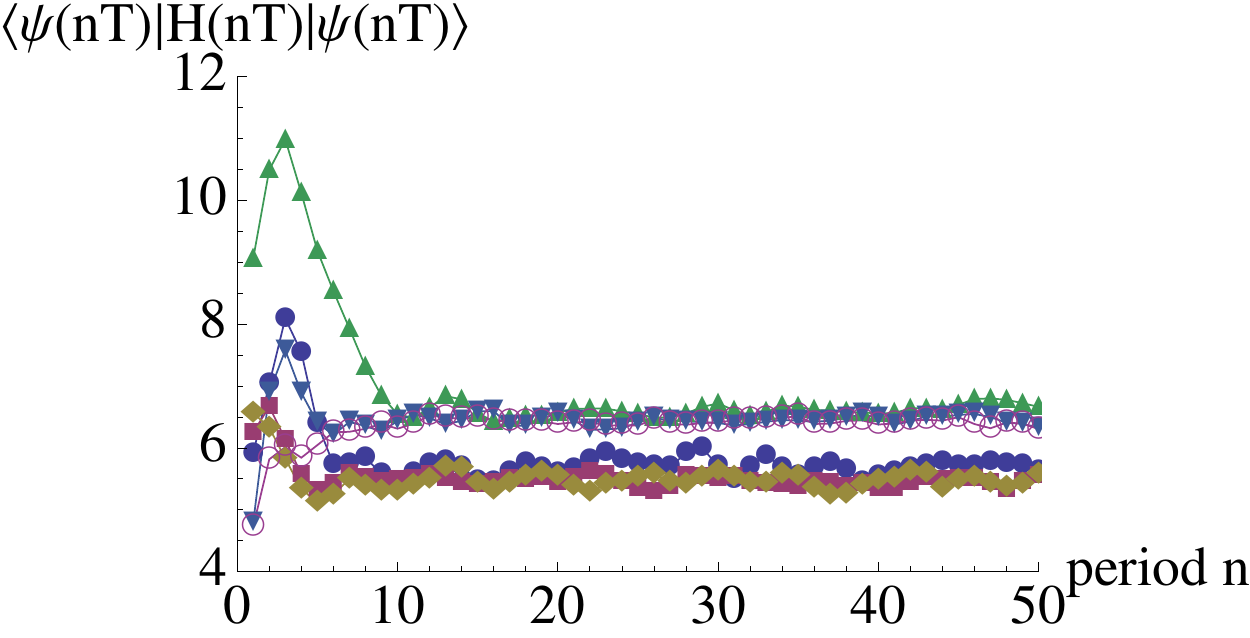}
\includegraphics[width=0.7\linewidth]{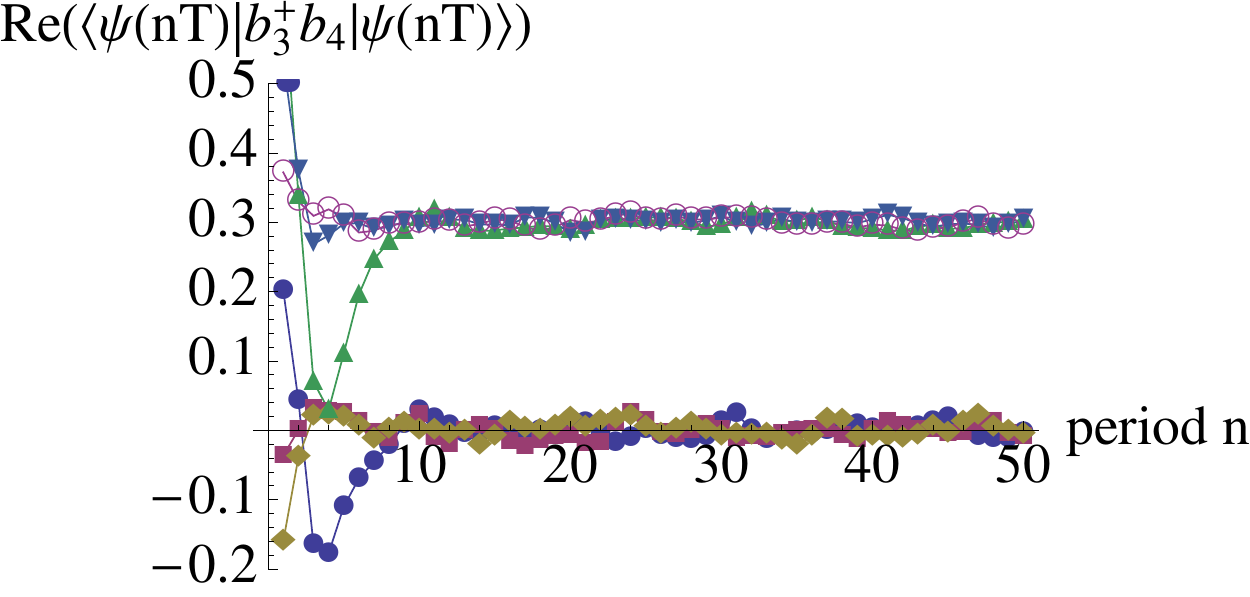}
\caption{\textbf{Top}: Time evolution of the Hamiltonian $H(t)$ 
(Eq.~\ref{H_Generic}) at the beginning of each period. 
The blue, gold red and gold data
points (three lower lines) correspond to different states, selected
from different parts of the band of a Hamiltonian (which is different
from the Hamiltonian used during the driving) and for for $L=12$,
$N=6$ so $D_{H}=924$. The three top lines show time evolution of
the same states, but for for $L=14$, $N=7$ so $D_{H}=3003$. In
both cases, $u/\omega=5$. \textbf{Bottom}: Similar results are
obtained for other correlators, such the real part of as 
$\la b_{3}^{\dagger}b_{4} \ra$. The
bottom three lines are for $L=12$, $N=6$ while the top three lines
are for $L=14$, $N=7$, and are offset vertically for clarity, but actually 
they also oscillate about $0$. (Taken from~\cite{LDM_PRE}). 
}
\label{fig:dynamical-H}
\end{figure}
Suppose a generic interacting non-integrable many-body system without disorder and localization is subjected to moderate periodic drive (i.e.,
the drive amplitude and frequency are comparable with the other 
couplings in $H(t)$). In such a system, the energy is not conserved, and since there is no underlying integrability, there is no obvious (periodic) conserved quantities to constraint the dynamics. This implies, 
the system will keep on absorbing energy without bound, and will asymptotically approach a state which looks locally like an infinite-temperature state. 
\\  

This picture is consistent with our common intuition about a closed Floquet system which does not dissipate 
out its energy, as reflected, say by Fermi's Golden rule for periodically
perturbed quantum systems with continuous spectrum (see, e.g.~\cite{Sakurai}). For systems with finite spectral width, the energy absorption will stop, but 
for any generic initial state, the final state will be a random superposition
of all states in any local basis. This means, {\it each} Floquet eigenstate
must be thermalized to infinite temperature! This Floquet ETH picture was proposed and verified for generic Hamiltonian irrespective of the form of the drive as long as it is moderate~\cite{Alessio_Rigol_PRX,LDM_PRE,Ergodic_NonErgodic_Heating_Abanin}. 
\\

Of course, one should keep in mind that even a 
quantum-chaotic time-periodic Hamiltonian $H(t)$
can have symmetries which it respects at all time, hence $\heff$ should also
respect those symmetries. In such a case, the chaotic hybridization of states due to the drive will hence happen only within each such symmetry sector, and not across them. Obviously, in that case, 
the quantities whose conservation are protected by the symmetry will not 
exhibit Floquet thermalization.\\

\begin{figure*}[t!]
\centering
\includegraphics[width=0.32\linewidth]{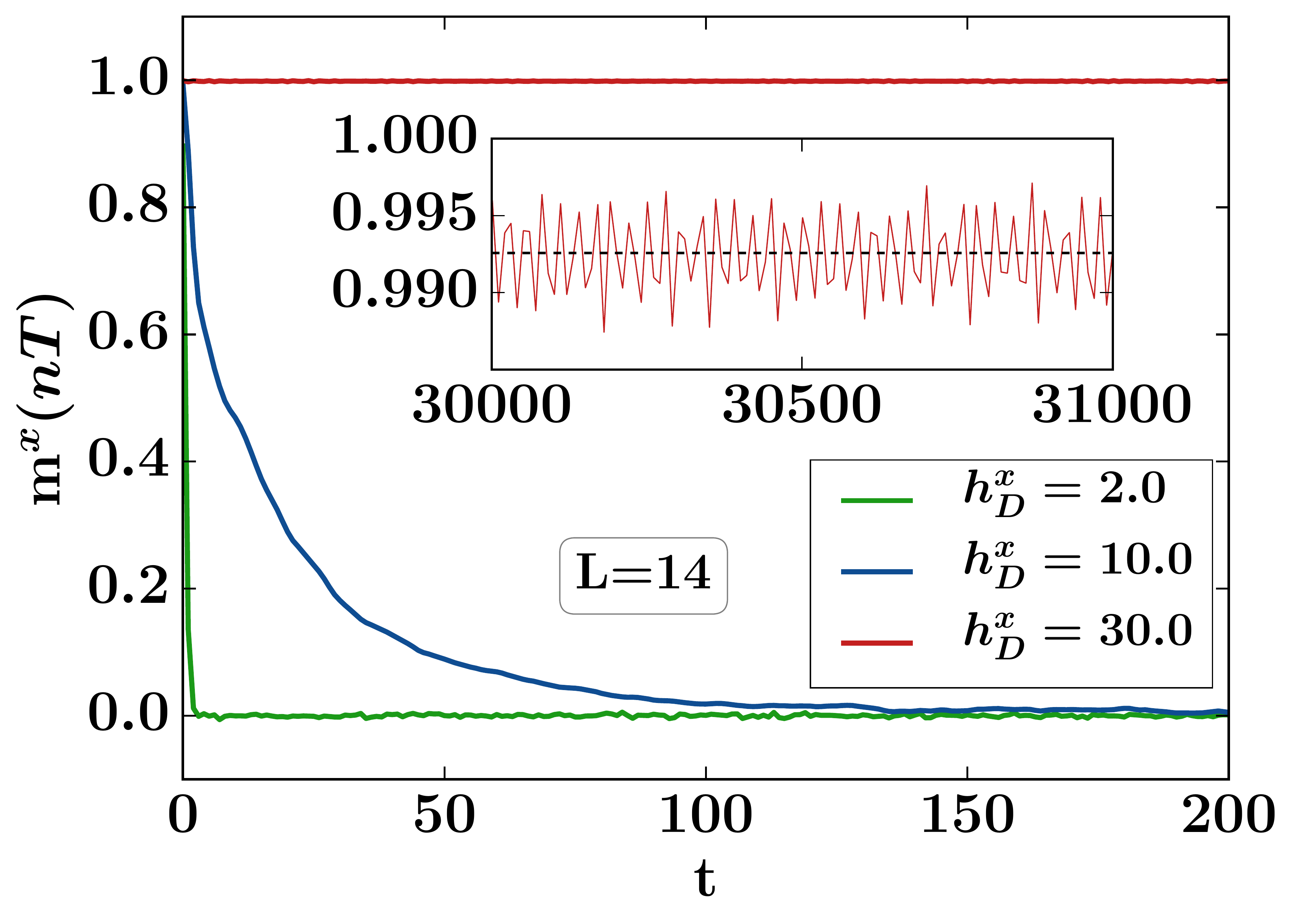}
\includegraphics[width=0.32\linewidth]{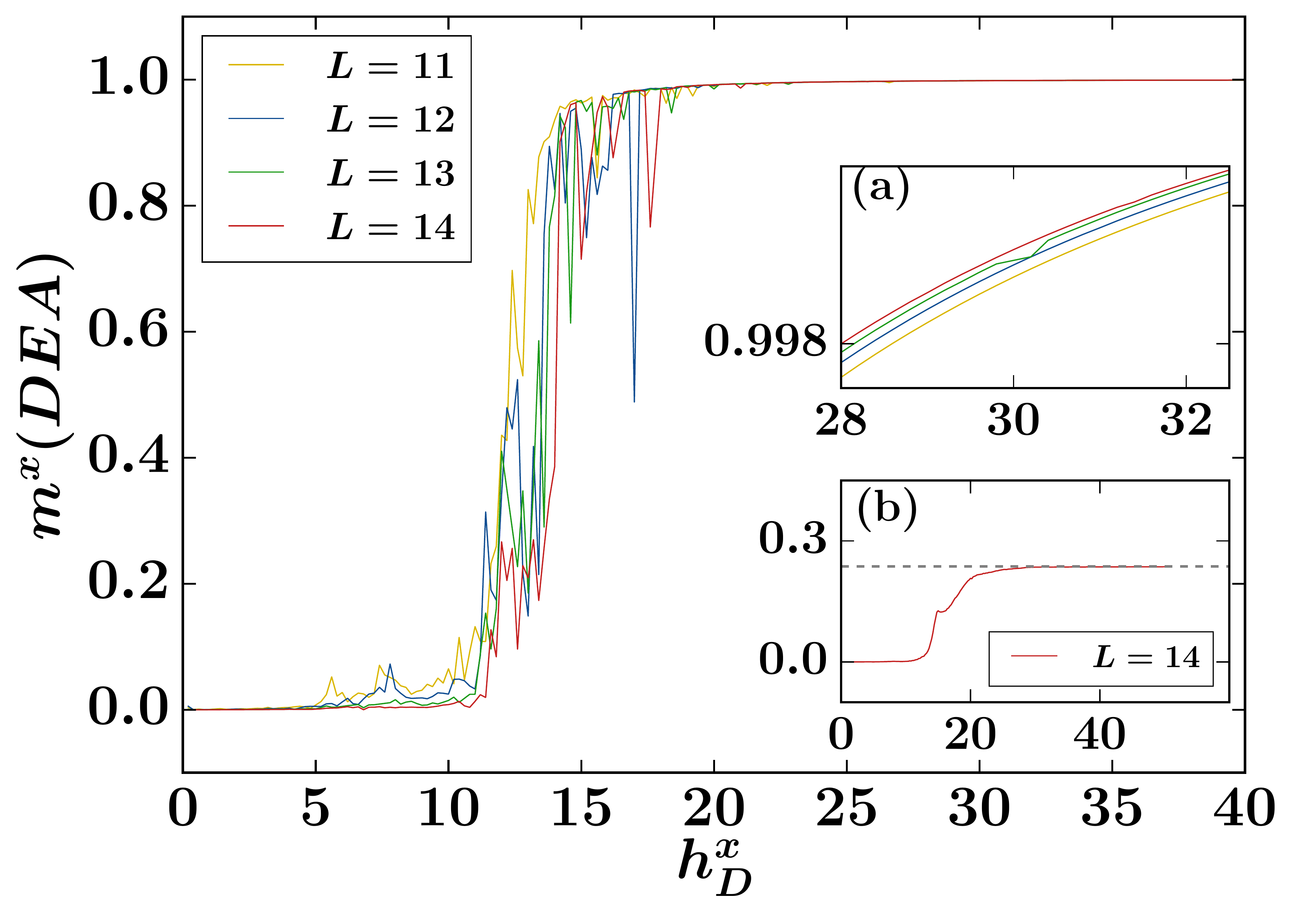}
\includegraphics[width=0.32\linewidth]{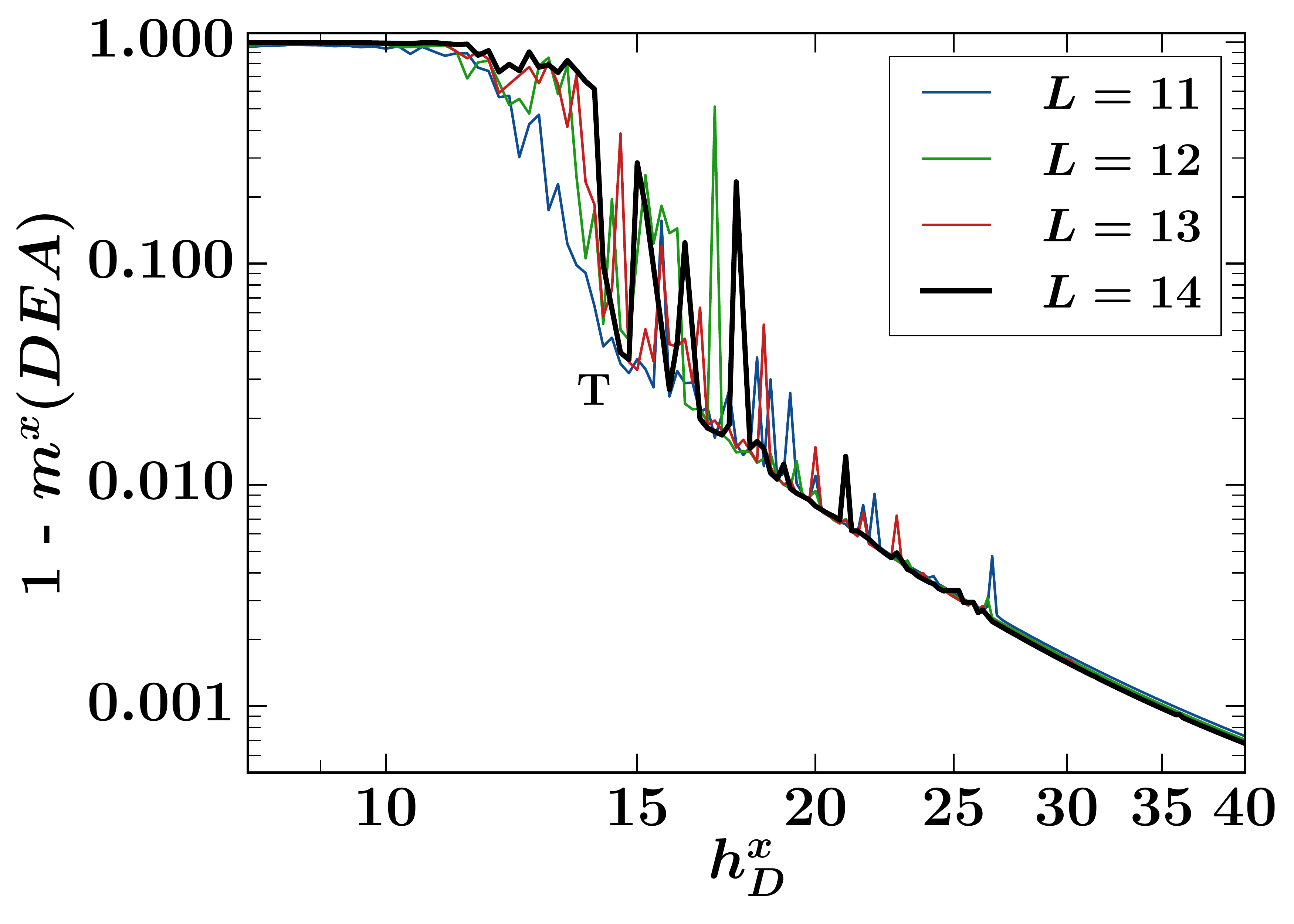}
\caption{
Freezing and its threshold.
{\it Left frame}: $m^x(\ell~T)$ for different driving
strengths showing initial state memory for strong driving. The inset zooms in on the long-time behaviour; the black 
horizontal line denotes the DEA of the magnetisation. 
{\it Middle frame}: Remnant magnetisation
as a function of driving strength for different system sizes. The 
high-field regime (top inset) shows an {\it increase} of the remnant magnetisation with $L$. The bottom
inset shows the diagonal ensemble average (DEA) of $m^x$ vs.\  drive amplitude for a `generic' state whose net 
initial magnetization is marked with the horizontal line, which remains almost unchanged for very strong drives. 
{\it Right frame}: Same data as middle frame on a doubly logarithmic plot for $1 - m^{x}{\rm (DEA)}$ .  The deviation away from 
almost complete thermalization gets steeper and moves towards the right with increasing system size. The curves
appear to accumulate from the left at a `threshold point' ({\bf T}) which itself appears to move little as the system size
is increased from $L=11$ to $L=14$.
(Taken from ~\cite{Onset}).
}
\label{fig1:Dynamics_mxDEA}
\end{figure*}
\noi
For illustration, in~\cite{LDM_PRE},  
a non-integrable model of hardcore bosons on a lattice given by the Hamiltonian 
\bea 
H(t) &=&-\frac{1}{2}\sum_{i}b_{i}^{\dagger}b_{i+1}+hc+V_{1}\sum_{i}n_{i}n_{i+1} \non \\
&+& V_{2}\sum_{i}n_{i}n_{i+2}+u\sum_{i}V_{i}(t)n_{i}
\label{H_Generic} 
\eea 
\noi has been considered,
with a time-periodic potential $V_{i}(t)=\widetilde{u}\left(t\right)\left(-1\right)^{i}$
with $\widetilde{u}(t)=+1$ for $0<t<T/2$ and $\widetilde{u}(t)=-1$
for $T/2\leq t\leq T$ ($J=V_{1}=V_{2}=1$). 
In particular, taking the above Hamiltonian as an example, 
it was shown that expectation value (eigenstate expectation values or EEV) of a 
local operator over all the Floquet eigenstates $|\al_{\epsilon}\rangle$
are almost same, with negligible state-to-state fluctuations across the spectrum (Fig.~\ref{fig:eev-example}). 
It shows that EEV for every Floquet state is almost same (a very narrow distribution., whose width was shown to vanish with 
increasing $L$), and are almost equal to the infinite temperature average.\\

The consequence of this on the late-time behaviour was also demonstrated in~\cite{LDM_PRE}. For a 
generic initial state $|\psi(0)\ra = \sum f_{i} |\al_{\epsilon}\rangle,$
the late-time expectation value of a local observable $\Ob$ will be given by
Eq.~\ref{Flq_DE_O}. Now according to Floquet ETH, that for every $\al,$ we have
\beq 
\la\al_{\epsilon}|\Ob|\al_{\epsilon}\ra \approx \la\Ob \ra_{\infty},
\eeq 
\noi where $\la\Ob \ra_{\infty}$ is the infinite-temperature average. Then, Eq~\ref{Flq_DE_O} immediately gives,
\beq 
\lim_{\ell\to\infty}\la\psi(\ell~T)|\Ob|\psi(\ell~T)\ra \approx \la\Ob \ra_{\infty},
\eeq 
\noi regardless of the specifics of the initial state and also the 
details of the Hamiltonian (since Floquet ETH is expected to hold for 
any generic Hamiltonian). This is demonstrated
in Fig.~\ref{fig:dynamical-H} for the Hamiltonian in Eq.~\ref{H_Generic}.\\

\noi 
{\bf Floquet ETH and $\heff$:} 
Floquet ETH will require $H_{eff}$ to be a non-local operator --
otherwise $H_{eff}$ itself will serve as a stroboscopically conserved local
operator, and the local description will be a Gibbs' ensemble 
with a $H_{eff}$ serving as a local Hamiltonian, rather than an
infinite temperature ensemble. Any series expansion of $\heff$ in terms
of local operators hence must not converge. 
Else one can obtain a controlled approximation for $\heff,$ and using that, an approximate Gibbs' distribution. This, of course, is a necessary but not sufficient 
condition for an unbounded Floquet heating.\\

\noi
{\bf Floquet Prethermalization}
We conclude this section briefly mentioning an
ephemeral but experimentally relevant phase 
of Floquet matter, namely, the prethermal phase.
It was shown that the phenomenon of prethermalization (see, e.g.,~\cite{Prthrm_1,Prthm_2,Prthrm_3,Prthrml_4}), i.e., existence
of a substantial time interval $t^{*}$ needed by an interacting, non-integrable
many-body system to exhibit the chaotic dynamics in its relaxation behaviour starting from an atypical initial state, also persists for Floquet systems~\cite{Eckstein_Canovi_Flq_Prethrm, Rigol_II}. In the early settings, prethermalization was shown for weakly interacting models which are in proximity with integrable ones: after switching on the integrability breaking interaction to an integrable Floquet system, the dynamics of the system continues respecting the periodic conservation laws (and the PGE) pertaining to the integrable system till some time $t^{*}.$
After that, the system thermalizes steadily. Later it was shown, prethermalization does not necessarily occur only in systems close to
integrability, but can exhibit memories of an atypical initial state for a long characteristic time in any generic system under fast drive~\cite{Kuwahara_Mori_Saito,Mori_Kuwahar_Saito_PRL,Dima_Floquet_Prethermalization,Flq_Prethermal_Wiedinger_Knap,Flq_Prthrm_Machado_Yao,LMSK,Bukov_Flq_Prethermal}, and
even in some slowly driven systems~\cite{DeRoeck_Verreet_Low_Freq_Prethrm}.
 Such prethermal Floquet states can host interesting
 though metastable phases of quantum matter, and can often be described by a PGE~\cite{Eckstein_Canovi_Flq_Prethrm}, or a thermal (Gibbs) ensemble~\cite{Flq_Prthrm_Machado_Yao}, or 
 even some interesting non-thermal phases~\cite{LMSK, DeRoeck_Prethrm_No_T}. 
 These phases can be of considerable 
 practical interest from a viewpoint of contemporary experiments -- perfect quantum dynamics
 can be simulated only for a finite, rather short
 durations, after which several unavoidable imperfections that entail a real-life setup will
 anyway bring in decoherence. If the life-time of a prethermal phase is comparable with the experimentally allowed time-windows, it is as good as a stable phase from experimental perspective. 
 Of course, a long-lived prethermal
 phase also appears generically when the drive amplitude is below a threshold~\cite{Boris_Fine_low_hD_Heating_Suppression}.
 
 In the next section we will see, stable Floquet states are possible in clean, interacting non-integrable systems if the drive is
 strong enough.
\begin{figure*}[t!]
\begin{center}
\includegraphics[width=0.45\linewidth]{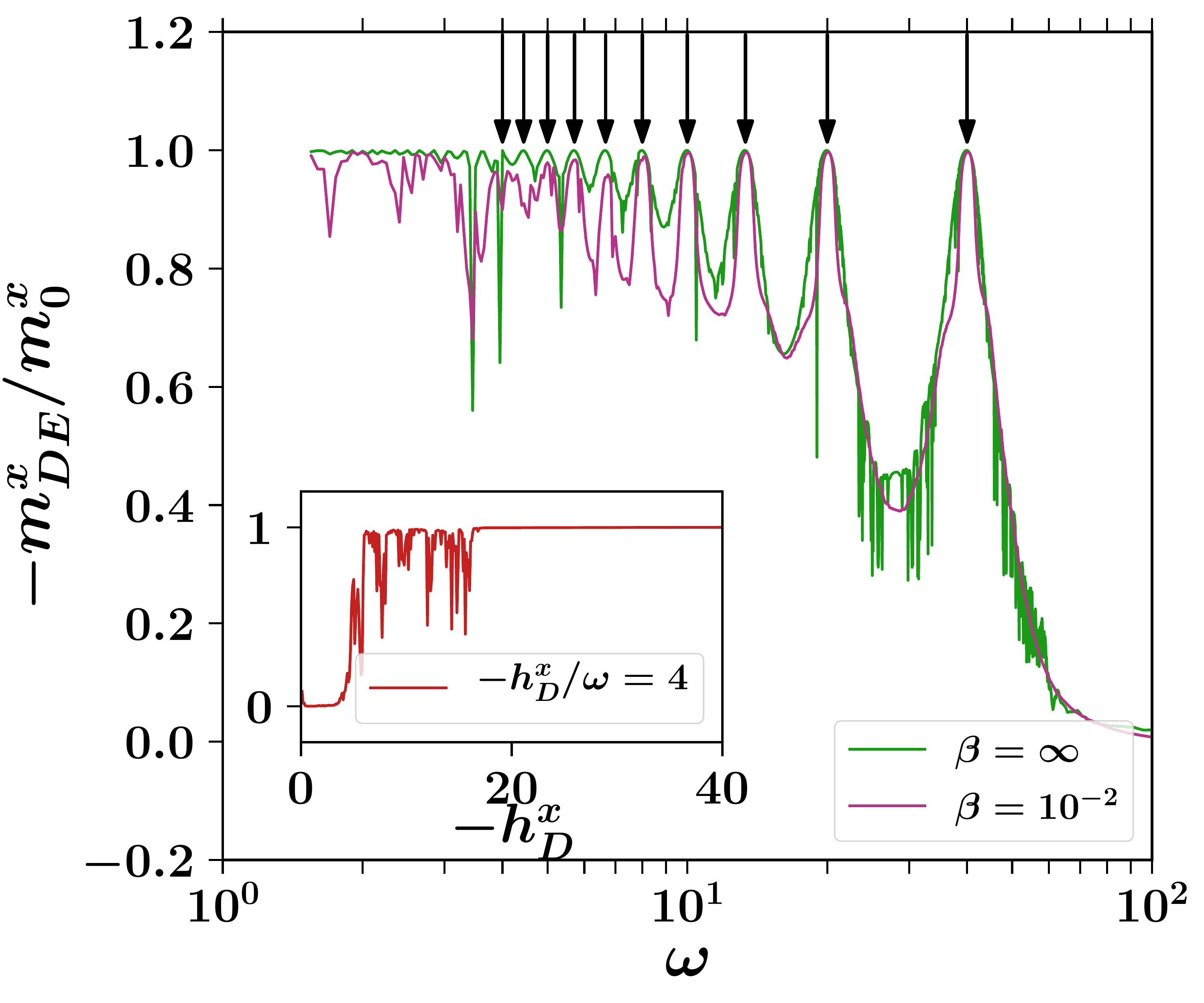}
\includegraphics[width=0.45\linewidth]{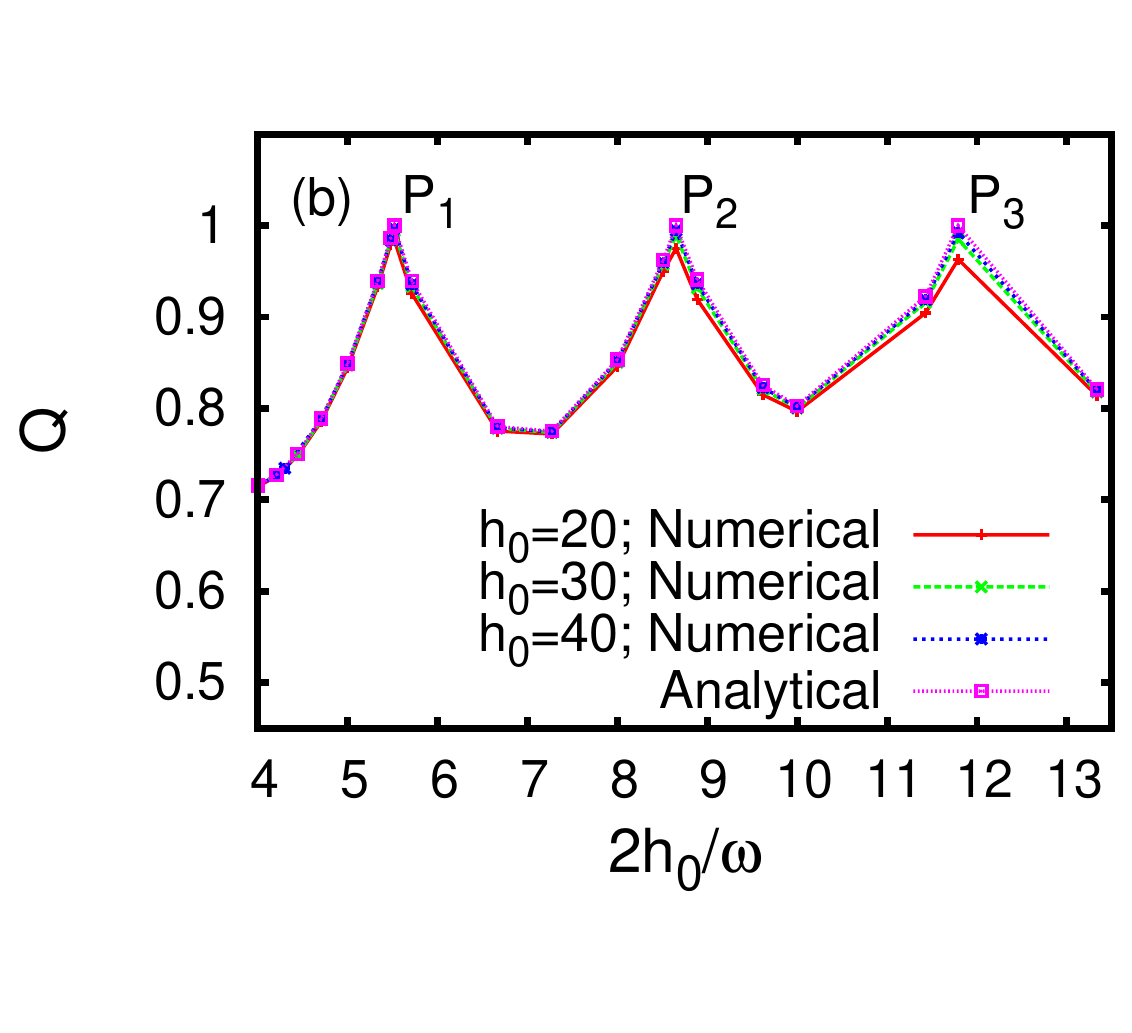}
\end{center}
\caption{Dynamical Many-body Freezing in {\bf (Left Frame):} non-integrable and (b) 
integrable Ising chains.
While the non-integrable freezing structure shows no perceptible debilitation
with increasing $L,$ the integrable structure is in the $L\to\infty$ limit.
	{\bf(Left frame):} $m_{DE}^{x}/m^x_0$, the ratio of magnetizations after. 
infinite (diagonal ensemble average) and 0 (initial state)
cycles versus drive frequency $\om$. Freezing, reflected in a large value of this ratio, occurs over a broad range of $\om$, and is strongest at 
particular {\it freezing points} (marked with arrows) $h^{x}_{D} = k\om$ (Eq.~\ref{Frz_Cond_Interacting}), where $k$ 
is an integer (for $h^{x}_{D }= - 40$ here, the arrows mark $\omega 
= 40/k;~ k=1,2,..,10$).
Results are shown for zero and high-temperature initial states: the former the 
ground state of $H(0)$ (which gives an initial magnetization $m^{x}(0) \approx 1$), 
and the latter the Gibbs state with $\beta = 10^{-2}$ 
($m^{x}(0) \approx 0.05$) for $H_{I}$ of the form $H(0)$, 
but with $h_{D}^{x}=5$; all 
other parameters are the same as the driven Hamiltonian, namely, $J=1, ~\kappa 
= 0.7\pi /3, ~h^{x}_{0}= e/ 10, ~h_{D}^{x}=40, ~h^{z}=1.2, ~L=14$. The sharp 
dips in the green lines
represent resonances, discussed in detail in the main text on Floquet-Dyson 
perturbation theory. Parameters are chosen to avoid these resonances. The {\bf inset} shows $m^x$ as a function of $h^{x}_{D}$ for a fixed ratio $|\hd/\om| = 4,$ showing the threshold  ($h^{x}_{D} \approx 20)$ above which the freezing sets in.
{\bf(Right frame):} Result for infinite-time average magnetization $Q = \overline{m^{z}}$ for an integrable chain (Eq. ~\ref{H-Ising}). 
The analytical result (Eq.~\ref{Q_Analytic}) is in the $L\to\infty$ limit. 
Left and right frames are from Refs.~\cite{Asmi_PRX} and~\cite{AD-DMF} respectively. 
}
\label{Peak_Valley_Steps} 
\end{figure*}
\section{Strongly Driven Clean Interacting Floquet Matter and
The Absence of Unbounded Heating}
\subsection{Floquet Thermalization Threshold}
The locally infinite-temperature like scenario that 
pertains to Floquet-ETH in interacting systems under moderate drive,
undergoes a sharp qualitative change as the drive amplitude is increased. 
Beyond a sharp threshold value of the drive amplitude, it shows a strong
memory of its initial state~\cite{Onset}.
The model taken for illustration is given by the $T-$periodic Hamiltonian
\begin{equation}
H(t) = H_{0}^{x} + H^{z}_{0} + \Sgn\left[\cos{(\omega t)}\right]\, H_{D},
	\label{Eq:Ham1_1}
\end{equation}
with 
\begin{eqnarray}
H_{0}^{x} &=& -J\sum_{i=1}^{L}\sigma_{i}^{x}\sigma_{i+1}^{x}  +\kappa\sum_{i=1}^{L}\sigma_{i}^{x}\sigma_{i+2}^{x} -h_{0}^{x}\sum_{i}^{L}\sigma_{i}^{x}\non\\
H_{0}^{z} &=& -h^{z}\sum_{i}^{L}\sigma_{i}^{z}
\nonumber \\ 
H_{D} &=& -h_{D}^{x}\sum_{i}^{L}\sigma_{i}^{x}. 
	\label{Eq:Ham1_2}
\end{eqnarray}
\noi 
The $\sigma_{i}^\alpha$s are Pauli matrices. The chain has periodic boundary condition, but 
the translational invariance is tampered with, by putting $J_{L,1} = 1.2J$ 
and $\kappa_{L-1,1}=1.2\kappa.$  
This choice was to break all obvious symmetries.
In presence of the transverse field, the
the static part $H_{0} = H_{0}^{x} + H_{0}^{z}$
of the Hamiltonian
is known to be ergodic due to the four-fermionic interaction terms arising from the next-nearest neighbour interactions under the spin to fermion mapping, and also due to the longitudinal field.
Under the periodic drive with the amplitude $h_{D}^{x}$ above a sharp threshold, the system shows
strong freezing, while below the threshold it's behaviour is consistent with Floquet thermalization to a locally infinite temperature state. This has been shown in Fig.~\ref{fig1:Dynamics_mxDEA}. 
Interestingly, as emphasized in the rightmost frame of the Fig., the threshold value shows no perceptible dependence on the system-size $L.$ The infinite drive-time limit ($\ell~T,$ as $\ell\to\infty$) is captured by taking the diagonal ensemble average in the Floquet basis (see, Eq.~\ref{Flq_DE_O}).
\\

\subsection{Dynamical Many-body Freezing and The Emergent Conservation Law in Clean 
Floquet Matter}
\begin{figure}[h!]
\centering
\includegraphics[width=0.85\linewidth]{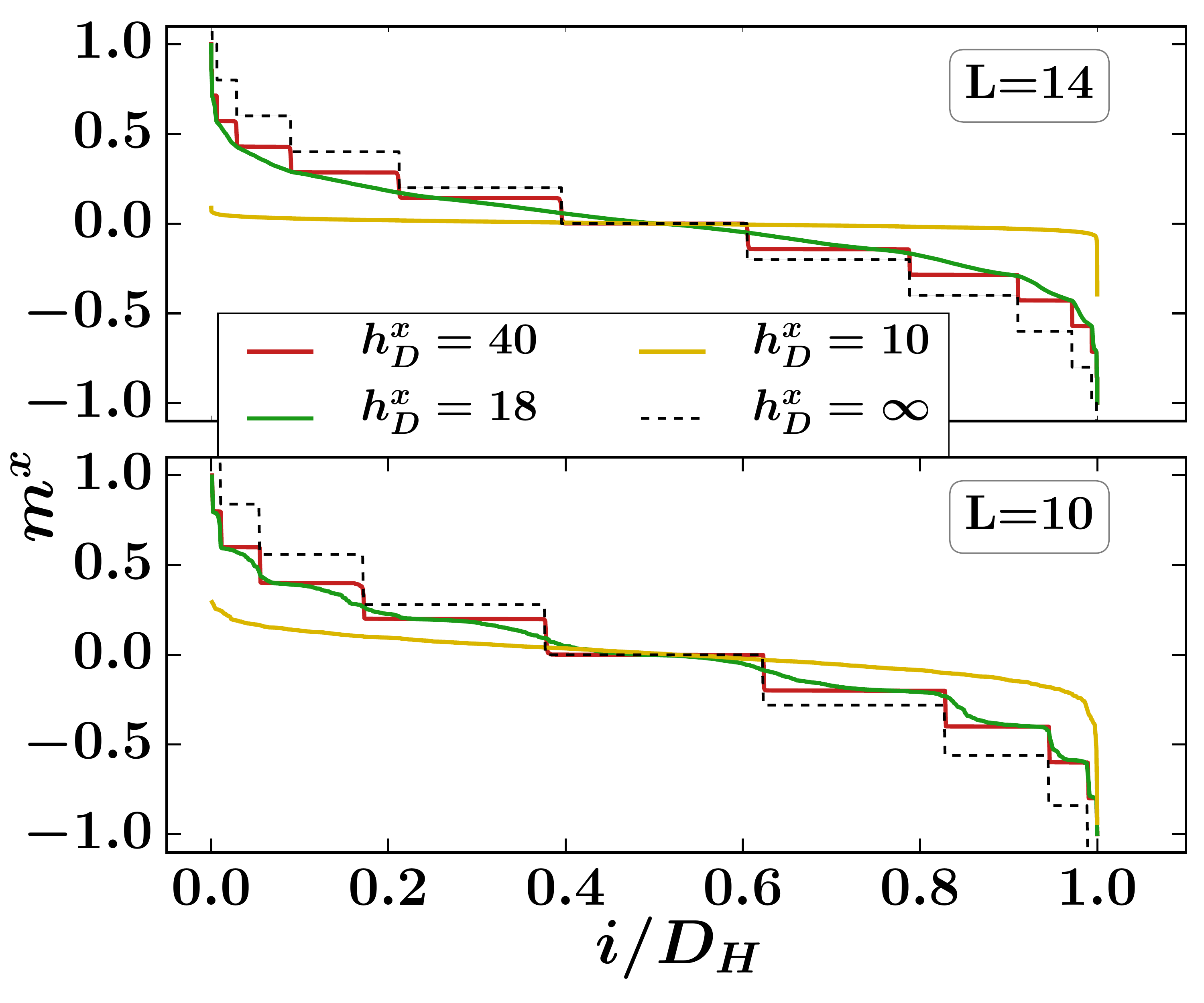}
\caption{
{Emergent conservation law for strong drives, as reflected in the Floquet eigenstates $|\mu_i\rangle$.} The expectation values of $m^x$ over the Floquet eigenstates  
arranged in decreasing order, for different values of $h_{D}^{x}$. 
Black dotted lines  ($h_{D}^{x} = \infty$) show the values of $m^x$ of the $x-$basis states (multiplied by a factor of 1.4 for visibility). 
	For $h_{D}^{x} = 40,$ clear step-like structures appear, indistinguishable from the 
	steps of $m^x$ for $x-$basis states for both system sizes $L=10,14$ (see Suppl. Mat. for finer details
	of $L$ dependence of this matching). For a lower drive value $h^{x}_{D}=18$, close to the threshold, 
	the curve smooths out, indicating weakening of the quasi-conservation, yet highly polarized Floquet 
	states are still substantial in number. For still lower values (e.g. $h^{x}_{D} = 10$), the curve finally flattens. 
	The pronounced asymmetry in the Floquet magnetization for lower values of $h_{D}^{x}$ is due to the small asymmetry in the drive. (Taken from \cite{Onset}).
	}
\label{Flq_IPR}
\end{figure}
\begin{figure*}[ht!]
\begin{center}
\includegraphics[width=0.85\linewidth]{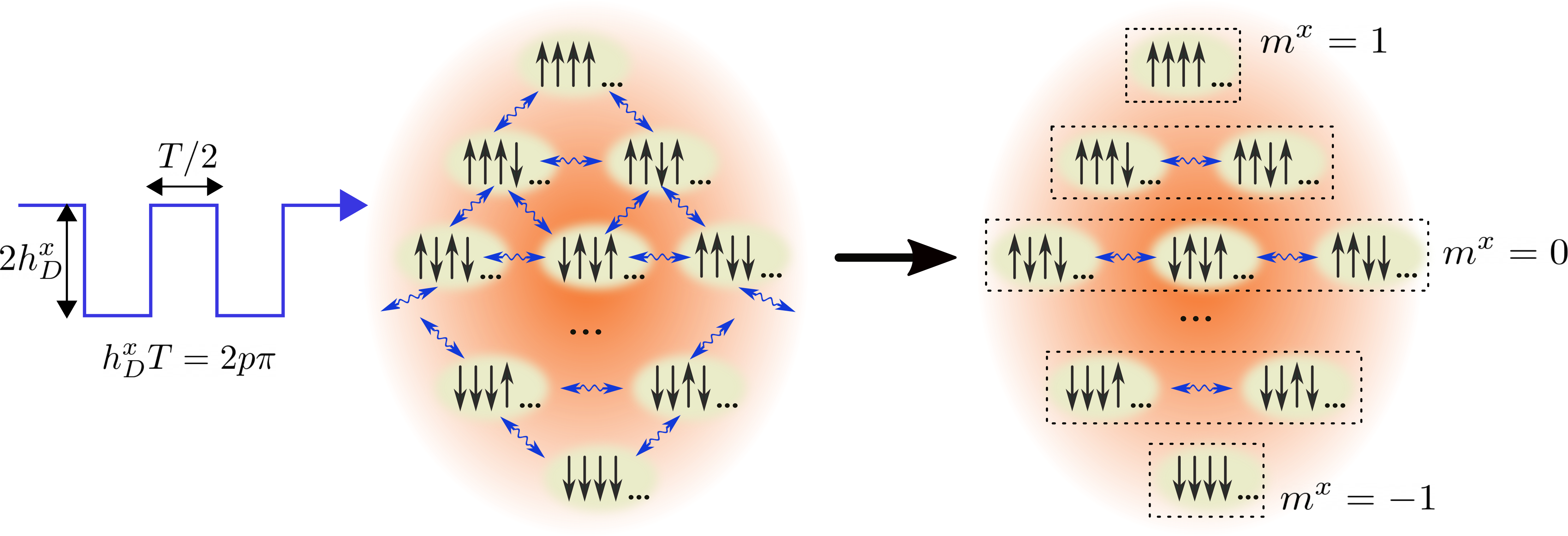}
\end{center}
\caption{
Fracturing of Hilbert space into dynamically disjoint sectors due to emergence of a local conserved quantity (LCQ) in a quantum chaotic many-body system. The condition $h_{D}^{x} = 2p\pi$ (Eq.~\ref{Frz_Cond_Interacting}) is the condition for maximal freezing 
of the magnetization (Fig.~\ref{Peak_Valley_Steps}), or, in other words, the condition under which the emergent conservation law is most accurate (see, \cite{Asmi_PRX}).
} 
\label{Scar_Cartoon} 
\end{figure*}

{\bf Dynamical Freezing:}
It was shown that the absence of thermalization and strong freezing is actually a manifestation of an {\it emergent} conservation law. Just
as it is in case of integrable Ising chain~\cite{AD-DMF}, here the magnetization $m^{x}$ emerges as an approximately conserved quantity under the drive with strength beyond the threshold. The conservation
(freezing in the language of Ref.~\cite{AD-DMF} -- see Sec.~\ref{DMF_Integrable}) becomes perfect (peak freezing) under the
drive condition
\beq 
 \hd = p~\omega, ~ {\rm or,} ~ h_{D}^{x}T = 2p\pi.
 \label{Frz_Cond_Interacting}
 \eeq 
 \noi ($p$ is an integer)
 as can be seen from the exact numerical results (Fig.~\ref{Peak_Valley_Steps}, left frame) for natural 
 (fully polarized) and generic (high temperature) initial states. This is also obtained from
 various analytical methods outlined in the following sections.
 This peak freezing
 condition is identical to that obtained for integrable Ising chain under square drive~\cite{SB_AD_SDG_2012} (Eq.~\ref{Frz_Cond_Sqr_Drv}).\\
 
{\bf Nature of the Conservation Law and Sub-system Entropy}
This can be gleaned from the expectation value of $m^x$ over the Floquet eigenstates $|\mu_{i}\ra,$ arranged in decreasing order of their value, against their label $i$ (in the same order) normalized by the Hilbert-space dimension $D_{H},$ 
as shown in Fig.~\ref{Flq_IPR}. These are compared for various values of the drive amplitude
$h_{D}^{x}$ against the step-structure obtained by similarly arranging the eigenvalue of $m^x$ for its unentangled eigenstates - we call them the $x-$basis states (these are the simultaneous eigenstates of all $\sigma_{i}^{x}.$)
The result show, for $h_{D}^{x} = 40,$ which is above the threshold (the threshold is $\approx 20,$ see Fig.~\ref{fig1:Dynamics_mxDEA}(c)), the magnetization of the Floquet eigenstates are almost indistinguishable from the magnetization of the eigenstates 
of $m^x$ in a one-to-one correspondence, which means that the Floquet eigenstates are at most hybridization of the $x-$basis states with the same eigenvalues for $m^x$, but not a superposition of them with different eigenvalues of $m^x$. In other words, each Floquet state is an eigenstates of $m^x,$ though they are free to delocalize within a given
eigen subspace of $m^x.$ The Hilbert space is thus fractured into disjoint eigen subspaces of $m^x,$ and it emerges as a conserved quantity (Fig.\ref{Scar_Cartoon}). As $h_{D}^{x}$ is reduced and the
threshold is approached, the step structure smooths out, indicating hybridization of $x-$basis states ($h_D^{x}= 18$), and below the threshold ($h_{D}^{x} = 10$) it flattens out completely to zero, indicating all the Floquet states are locally thermalized to infinite temperature. \\

Whether conservation of $m^x$ alone suffices to determine the ensemble, i.e.
whether $\rho_{_{sub}} = e^{-\lambda m^{x}}/Z$ is an accurate description 
of the Floquet ensemble is still an open question. In particular, this seems not to be the entire picture away from the peak freezing points. There $m^x$ is only approximately conserved. The constraints and their nature are yet to be identified. To a first approximation, the constraint consists of a local Hamiltonian given by the first few terms of the moving frame Magnus expansion away from the freezing peak, assuming the expansion is asymptotic. Further investigation in this direction is in progress.
\\

The effect of the emergent conservation law on the growth of sub-system entropy
has been illustrated in Ref.~\cite{Asmi_PRX}. The evolution of the half-chain entanglement was tracked, taking various $x-$basis states belonging to different eigen-subspaces of $m^x$ of various sizes as the initial states. The entropy growth reflects the nature of the conservation - it is not that 
the system remains close to an $x-$basis state because $h_{D}^x$ is large compared to other scales in the problem. In fact, most of the $x-$basis states will evolve substantially to generate extensive
sub-system entropy, while some will not. What is respected, is the
{\it global} conservation of $m^x:$
the dynamics spreads an $x-$basis state only within its own eigen-subspace corresponding to the same eigenvalue of $m^x.$
For example, a fully $+x-$polarized state is the only member of the subspace with $m^x=1$, hence taking this as a initial state shows no entanglement growth even after million drive cycles.  But taking any $x-$basis state with $m^x=0$ (e.g., a N\'{e}el ordered state) results in substantial growth of the entanglement, since the $m^{x} = 0$ sector has several members over which the evolving state can delocalize.
\\

\subsection{
Stability of the Conservation Laws and the Thermodynamic Limit
}
Stable Floquet states have been observed numerically under moderate drive 
in various clean interacting finite-size systems~\cite{Adhip_Diptiman,Sayak_Utsa_Amit,
Qin_Hofstetter,gil-fss,Prosen_prl_98,Prosen_Tilted_NoHeat,AL_DL_PRB}. However, no evidence of stability was reported for those moderately driven systems in the large $L$ limit (see, also \cite{Achilleas_David_Infinitesimal_Heating}). 
As already mentioned above, the situation changes drastically if the drive amplitude crosses a threshold strength~\cite{Onset}. A systematic  finite-size analysis of the stability of $m^X$ was made for various interacting non-integrable models - Ising and Heisenberg spins with interactions of various ranges and locality (including three spins interactions). None of them shows any discernible decline in the conservation with the system-size~\cite{Asmi_PRX}. Moreover, 
for a large class of models (Ising and Heisenberg interactions of virtually any kind and in any lattice dimension in the static Hamiltonian),
it was shown that a moving frame Magnus expansion for $\heff$ (applies independent of $L,$ hence holds as $L\to\infty$) is an asymptotic expansion, and it is consistent with the stability 
 of $m^x$ observed from the finite size analysis of the exact numerical results. That is an indicative of the survival of the conservation in the $L\to\infty$ limit, though this is certainly not a conclusive proof of the same.\\

Finally, it is worth emphasizing that the stability of the emergent conserved quantities in the thermodynamic limit in non-interacting systems is far from
a trivial consequence of integrability. 
Let us consider the non-interacting Ising chain (\cite{AD-DMF,SB_AD_SDG_2012})
under strong periodic drive, where $m^x$ (the component of 
magnetization coupled to the driven field) emerges as a conserved quantity.
As discussed in Sec.~\ref{Sec:Floquet_PGE}, the late-time state of these
systems can be described by a PGE constrained by $L$ exact 
local conservation laws. The commutators of these LCQs
$\Ik$s with $m^x$ does not essentially vanish in general. Hence according to 
principle of unbiased distribution (entropy maximization subject only to the conservation of $\Ik$s) and Floquet ETH,  and a common 
eigenstate of all $\Ik$s is not also an eigenstate of $m^x$ in general, and hence the (stroboscopic) dynamics is not supposed to conserve $m^x.$ 
Nonetheless, $m^x$ emerges as an approximate conserved quantity
which is almost perfect under the freezing conditions, and it persist
stably the limit $L\to\infty$ at all time. Fig.~\ref{Peak_Valley_Steps} shows the
peak-valley structure of the late-time $m^{x}$ vs $\omega$ for the non-integrable and integrable Ising cases, the integrable one is calculated in the limit $L\to\infty.$ 

As discussed above, this stability
in the thermodynamic limit has no trivial connection with the integrability 
of the model. This can also be gleaned from the late time (PGE) behaviour
of sub-system entanglement entropy -- a state with entropy independent of system-size, like one dynamically frozen close to the fully polarized state is certainly not an essential consequence of integrability. In general, such states exhibit sub-system entanglement entropy that grows as $l_{sub}^{\nu},$ with $\nu >0 $, where $l_{sub}$ is the sub-system size (see, e.g., ~\cite{Sourav_Entanglement}). 
\\

\subsection{Emergent Symmetries in Clean Floquet Matter}

Floquet matter trivially inherits the symmetries present in $H(t)$
at all time. In addition, there can appear emergent symmetries. 
These symmetries can be approximate, in the sense, they are manifest in 
few initial terms in an asymptotic or convergent expansion of $\heff.$  Here we illustrate the
idea briefly with an example.\\

Let us consider the Floquet system described by the Hamiltonian of the form
\beq
H(t) = H_{0}^{x} + H_{0}^{yz} +  h_{D}^{x}~\Sgn[\cos{(\omega t)}]~H_{D}^{x},
\label{H_Gen}
\eeq 
\noi
where $H_{0}^{x}$ and $H_{D}^{x}$ depends only on $\sigma_{i}^{x}$ while $[H_{0}^{yz}, H_{0}^{x}] \ne 0,$ 
$[H_{0}^{yz}, H_{D}^{x}] \ne 0,$ and all of them are time-independent. An example is the Hamiltonian given by Eqs.~\ref{Eq:Ham1_1},~\ref{Eq:Ham1_2}. Direct Magnus expansion for $\heff$
is problematic for $H(t)$ of this form, since $h_{D}^{x}$ is large, and will appear in the 
numerator of the expansion killing the possibility of having even an asymptotic expansion for moderate values of $\omega,$ which is the regime of interest. A way around is to go to a time-dependent frame where there will
be no large term in the time-dependent Hamiltonian (see e.g.~\cite{Anatoli_Rev}). This can be done by employing a time-dependent rotation
via the Unitary
\beq 
W(t) = \exp{\left[-i\int_{0}^{t} dt^{\prime} ~\Sgn[\cos{(\omega t^{\prime})}])~H_D^{x}\right]}
\label{U_Mov}
\eeq 
\noi
Then one applies this to $H(t)$  (Eqs.~\ref{Eq:Ham1_1},~\ref{Eq:Ham1_2}), and does the Magnus expansion in the moving frame for the transformed Hamiltonian~\cite{Asmi_PRX}. This
give, under the freezing condition $h_{D}^{x} = p \omega$ (Eq.~\ref{Frz_Cond_Interacting}). The first two terms of the expansion is simply proportional to $H_{0}^{x}$ - a static term that commutes with $H_{D}^{x} = m^x.$ This gives an example of an emergent $U(1)$ symmetry -- to the two initial orders,  $\heff$ 
 is invariant under any rotation about the $x-$axis, though $H(t)$ is not, since it contains the $H_{0}^{yz}$ term. \\
 
 A bit more interesting manifestation of this emergent symmetry is observed for the driven Heisenberg system~\cite{Asmi_PRX}.
Consider a Floquet system of the form in Eq.~\ref{H_Gen} with
\bea 
H^{yz}_{0} &=& -\sum_{i,j}\Jyij\si_i^y\si_{j}^y -\sum_{i,j}\Jzij\si_i^z
\si_{j}^z -h^{z}\sum_{i}\sigma^{z}, \non \\
H^{x}_{0} &=& -\sum_{i,j}\Jxij\si_i^x \si_{j}^{x} + \ka \sum \si_i^x 
\si_{i+2}^x - h_0^x\sum \si_{i}^{x}. ~~~
\label{H_Heisenberg} 
\eea
\noi Switching to the moving frame via the transformation 
in Eq.~\ref{U_Mov}, and doing the Magnus expansion with the transformed Hamiltonian one gets at the first order
\beq
H_{eff}^{(0)} ~=~ 
H_0^{x} ~-~ \frac{1}{2}\sum_{i,j}(\Jyij + \Jzij)\left[ \si_i^y 
\si_{j}^y + \si_i^z \si_{j}^z \right]
\label{Emerrgent_U1}
\eeq 
\noi
under the freezing condition $h_{D}^{x} = n\om.$
Though this term contains not only $H_{0}^{x}$ but also terms involving 
$\sigma^{y}_{i},\sigma_{i}^{z},$ it still respects the emergent
$U(1)$ symmetry and the associated conservation of $m^x.$ Surprisingly, the next order term is 
even more simple -- we just get~\cite{Asmi_PRX}
$$
H_{eff}^{(1)} = H_{0}^{x}.
$$
\noi 
In Ref.~\cite{Asmi_PRX}, there are strong numerical evidences of this freezing 
(and its system-size independence) for a plethora of interacting models falling within the remits of the above
analytical results. These complementary results strongly point towards the asymptotic nature of the 
expansion and consequent stability of the emergent conservation law in large systems.
Here, one must note that the above calculation has been explicitly done with the Floquet gauge set to $\epsilon = 0.$
\\

The moving frame thus beautifully betrays the emergent $U(1)$ symmetry (of course, confirmed up to the 1st order) due to the strong field in the $x-$direction.
 It is interesting to note that a usual (static frame) Magnus expansion would not be able to capture this. The first order 
 term $\heff^{(0)}$ will simply be the time-average of $H(t)$ 
 over a period, which will leave a strongly non-integrable Hamiltonian without the symmetry or the conservation. \\

\subsection{Heating and Resonances: The Floquet-Dyson Perturbation Theory (FDPT)}
Like any quantum many-body systems, the stability of a localized state
can be threatened by resonant transfer of energy between the localized states~\cite{Dalibard_Goldman_Monik_Cooper_Resonance,Bukov_Polku_Huse,Prosen_Anatoli_Replica}. For conservation of $m^x,$ the localization needs to happens
in the $x-$basis within different eigen-subspaces of $m^x$. Hence
it is natural to construct a perturbation theory in $x-$basis, and 
investigate the effect of little perturbation on the states localized in 
the said basis. The key results are (a) this
turns out to be the right basis, i.e., for expansion in this basis the asymptotic results are consistent with the exact numerical results and (b) relevant resonances can be avoided over a stretch of parameter space.\\

From the expression of $H(t)$ (Eq.~\ref{H_Gen}) note that in the absence of $H_{0}^{yz},$ the Hamiltonian (including the drive) consists only of $\sigma_{i}^{x}$s, hence each $x-$basis state is a Floquet eigenstates.
Now, since $h_{D}^{x}$ is large, we can treat $H^{yz}_{0}$ as a perturbation. At first sight, this might look like a lot of constraint on
 $H(t),$ but actuality, the only constraint is, the drive part should be of the form $h_{D}(t)~H_{D},$ where $h_{D}$ is large compared to all other couplings and $H_{D}$ is time-independent. This need not be a function of
 $\sigma_{i}^{x}$s only in general. The rest is just grouping terms
 under two heads, and applies to any general Hamiltonian: $H_{0}^{x}$
 is the static part of the Hamiltonian that commutes with $H_{D}$ (hence, in general, it also need not be a function of
 $\sigma_{i}^{x}$s only), and $H_{0}^{yz}$ is the static part 
 that does not commute with $H_{D},$ and
 $x-$basis are simply the common eigen-basis of $H_{D}$ and $H_{0}^{x}$.\\

Renaming $H_{0} = H_{0}^{x} + h_{D}(t)~H_{D}$ and $V = H_{0}^{yz},$ 
the basis is chosen to be eigen-basis of $H_0(t)$, 
denoted by $| n \ra$,  These are the unperturbed Floquet states -- this is possible since the drive term commutes with itself at all time, being of the form $h_{D}(t)~H_{D}$,and hence the unperturbed Floquet states are time-independent. They satisfy
\beq 
H_0 (t) | n \ra ~=~ E_n (t) | n \ra, 
\label{eig1} 
\eeq
\noi
and $\la m | n \ra = \de_{mn}$. 

Now, it can be assumed without any loss of generality that $V$ is 
completely off-diagonal in the unperturbed basis 
(since we can always absorb its diagonal part, if any, in $H_{0}$). 
This gives
\beq \la n | V | n \ra ~=~ 0 \label{nvn} \eeq
for all $n$. 
We will now find solutions of the time-dependent Schr\"{o}dinger 
equation
\beq 
i \frac{\pa |\psi_n\ra}{\pa t} ~=~ H(t) |\psi_n (t)\ra, 
\label{sch1} 
\eeq
which satisfy 
\beq 
|\psi_n (T)\ra ~=~ e^{- i \mu_n} ~|\psi_n (0)\ra. 
\label{floeig1} 
\eeq
\begin{figure}[t!]
\begin{center}
\includegraphics[width=0.98\linewidth]{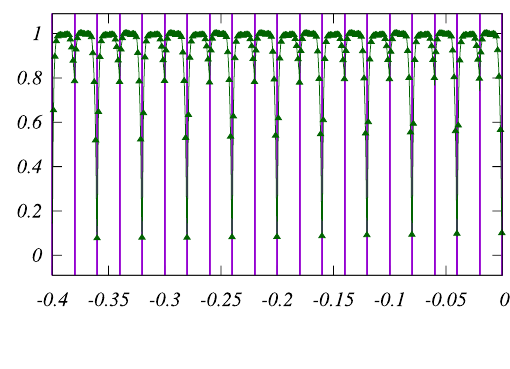}
\end{center}
\caption{
{\bf Freezing vs Resonances:} 
The plot shows magnetization ratio $m_{DE}^{x}/m^x_0$ versus $h_0^x,$, where $m^{x}_{DE}$ is the diagonal ensemble average, and $m^{x}_{0}$ is the initial magnetization.
The result is shown under a peak freezing condition given by $h_{D}^{x} = 40,$ and $\omega =0.04.$
The numerical results (green triangles) show, there is indeed strong freezing even at this very 
low frequency, but the freezing is drastically destroyed at well-separated points, around which the magnetization shows sharp dips. The vertical violate line marks the resonance points predicted by FDPT (Eq.~\ref{Gen_Res}). The Fig. shows that almost all the dips are predicted by FDPT. The
values of the other couplings are $J=1, ~\kappa = 0.7\pi/2, ~h^{x}_{0}= e/ 10, ~h^{z}=1.2, ~L=14.$
The numerical result is shown for a finite temperature density matrix with 
initial inverse temperature $\beta = 10^{-2},$ undergoing Schr\"{o}dinger evolution.
(Taken from~\cite{Asmi_PRX}).
\\
\label{Resonances} 
}
\end{figure}
For $V=0$, each eigenstate $|n\ra$ of $H_{0}(t)$ is a Floquet state, with 
Floquet quasienergy $\mu_{n}^{(0)} = \int_{0}^{T} dt E_{n}(t)$ (defined 
modulo $2\pi$).

For $V$ non-zero but small, a Dyson-like series is developed for the 
wave function to first order in $V$. 
Clearly $V$ is a small perturbation as long as $|V /h^{x}_{D}| \ll 1,$ though it 
can otherwise be comparable to or larger than the other couplings of the undriven Hamiltonian.
The $n$-th eigenstate can be manifestly written as a Floquet state
by inserting the time-dependent phase it gathers, and in terms 
of those, the Floquet state obtained by perturbing the unperturbed Floquet state $e^{-i\int_{0}^{t}E_{n}(t^{'}) dt}|n\ra$   
may be expressed as a linear combination of the other unperturbed Floquet states as
\beq 
|\psi_n (t)\ra ~=~ \sum_m ~c_m (t) ~e^{-i \int_0^t dt' E_m (t')} ~ |m \ra 
\label{nine}, 
\eeq
where $c_n (t) \simeq 1$ for all $t$ while $c_m (t)$ is of order $V$ (and
therefore small) for all $m \ne n$ and all $t$. \\

Then
substituting the form for the wave-function in Eq.~\ref{nine}
in the time-dependent Schr\"{o}dinger equation, one then demands this to be a Floquet eigenstates, i.e.,
\beq
|\psi_n(T)\ra = e^{-i\mu_{n}}|\psi_{n}(0)\rangle. 
\label{Floquet_Cond}\non
\eeq
\noi
Then taking
the overlaps with the basis states $|m\ra,$  one gets:
\beq c_m (0) ~=~ - i ~\la m | V | n \ra ~\frac{\int_0^T dt ~
e^{i \int_0^t dt' [E_m (t') - E_n (t')]}}{e^{i \int_0^T dt [E_m (t) - E_n (t)]}
~-~ 1}. 
\label{cmt2}
\eeq
\noi
 Note that $c_m (t)$ is indeed of order $V.$
 All these are well-defined provided the denominator on
the right hand side of Eq.~\eqref{cmt2} does not vanish. When it does,  we
hit a singularity under the condition
\beq 
e^{i \int_0^T dt [E_m (t) - E_n (t)]} ~=~ 1. 
\label{res1} 
\eeq
This is precisely the condition for a resonance
between states $| m \ra$ and $|n \ra$, and the above 
analysis breaks down. Now, if there are several states 
which are connected to $|n\ra$ by the perturbation $V$, Eq.~\eqref{cmt2} 
describes the amplitude to go to each of them from $|n\ra$.
Up to order $V^2$, the total probability of excitation away from $|n\ra$
is given by $\sum_{m \ne n} |c_m (0)|^2$ at time $t=0$. Further
simplification of Eq.~\ref{res1} gives the first order resonance condition
in the form
\beq
\int_{0}^{T} \left[E_{m}(t) - E_{n}(t) \right] dt = 2p\pi,
\label{Reso_Gen_Simple}
\eeq
\noi where $p$ is an integer including $0$, and $E_{m}(t)$ and $E_{n}(t)$ are
energies of two unperturbed states $|m\ra$ and $|n\ra$ respectively, differing from each other by a single spin flip. 
\\

Applying this to the specific case of Hamiltonian (Eqs.~\ref{Eq:Ham1_1}.~\ref{Eq:Ham1_2}), one gets the condition for
all first order resonances:
\beq 
h^{x}_{0}\si_{0} ~+~ J\si_{0}(\si_{-1} + \si_{1}) ~-~ \kappa\si_{0}(
\si_{-2} + \si_{2}) ~=~ \frac{p\om}{2}, 
\label{Gen_Res} 
\eeq 
\noi where we have used classical Ising variables
$\sigma_{0,-1,+1,-2,+2},$ each taking values $\pm 1,$ to compactly write the 32 possible different 
values of (unperturbed) energy change that might occur due to the flip of a single spin interacting
with its four neighbours in presence of a field on itself. The significance of those resonance points are immediately obvious once one looks into the numerical results (Fig.~\ref{Resonances}).
The predictions from the first order of FDPT (Eq.~\ref{Gen_Res})
accounts for all the sharp dips that indicates the failure of freezing and signature of unbounded heating. Finite size analysis (not shown) shows that these resonant drops grows deeper with increasing $L,$ while away from the resonances, $m^x_{DE}$ shows
no perceptible changes with $L.$ This indicates that to avoid heating, all one needs to avoid are those first order resonances. Of course FDPT predicts higher order resonances as well, but for large values of $\hd,$ somehow those higher order terms, along with their resonances, are suppressed. The 
first order resonances can always be spaced apart choosing the parameters properly, as can be seen from Eq.~\ref{Gen_Res}. For example, one can choose all the parameters to be integers except $\omega,$ which is chosen to be an irrational number. Then the resonances can be kept finitely apart even as $L\to\infty.$ Hence one would get finite patches on the parameter space, such that if the set of values of the parameters lie on the patch, then the Floquet matter will be free from first order resonances and hence the thermalization. It is interesting to note  that in the numerical studies, higher order resonances do not appear even in the infinite time limit ($\ell~T,$ as $\ell\to\infty$) captured by the diagonal ensemble average. \\

It is interesting to note, the FDPT also gives the peak 
freezing condition $h_{D}^{x} = k\omega$
(Eq.~\ref{Frz_Cond_Interacting}) for the Floquet
eigenstates obtained by perturbing the fully-polarized initial state~\cite{Asmi_PRX}. The perturbed Floquet eigenstate can be calculated to first order in this case, and 
expectation value of $m^x$ over it is given by
\bea 
1 ~-~ m^x 
\approx ~\left(\frac{h^z}{h_D^x} \right)^2 ~\frac{4 (1 ~-~ 
A^2 T^2/8)~ \sin^2 (h_D^x T/2)}{A^2 T^2}, \non
\eea
for $\omega \gg A,$ where $A = 4(J-\kappa) - 2h_{0}^{x}.$
From the above equation, 
one gets $m^x \approx 1$ under the
peak freezing condition $h_{D}^{x} = k\omega.$
\\

\subsection{Floquet Scars: The Statistically Relevant Ones}
In simple terms, a quantum many-body scar is a ETH violating high energy (finite energy density)
eigenstate of a quantum chaotic many-body Hamiltonian~\cite{Scar_Abanin_NatPhys,Scar_Lukin_Nature,
Scar_Abanin_NatPhys, Scar_Shiraishi-Mori, Scar_Abanin_PRB, Scar_Vedika_PRB, 
Scar_Lukin_PRL, Scar_Serbyn_PRL, Scar_Bernevig_1, Scar_Vedika_Rahul,Scar_Bernevig_2,Scar_Paul,Debu_Babun_Scar_Zero_Mode}
(see, however, Ref.~\cite{Rigol_Comment}). Such scars are usually outliers in the spectrum, occurring at zero measure. Hence their existence usually do not affect the statistical mechanics of their hosts.\\

In localized systems, periodic drive often fails to delocalize certain/all of the localized eigenstates, and they are often dubbed as Floquet scars. First example of those 
are actually the Floquet eigenstates of a Floquet MBL systems~\cite{AL_AD_RM_PRL_2014,Abanin_MBL}. 
Other interesting examples include (but are not limited to)
Stark localized systems under periodic drive~\cite{Bhaskar_Scar}, scars due to projectors in the Hamiltonian surviving periodic drive~\cite{Flq_Scar_Saito,Flq_Scar_Mizuta} etc. 
These, like the equilibrium scars, do not alter
the statistical mechanics of the system.
\\

A generic way of producing  statistically significant 
Floquet scars (i.e., Floquet eigenstates that violates Floquet ETH) in a  quantum-chaotic system, is to drive it periodically keeping the drive strength above the Floquet thermalization
threshold~\cite{Onset}. This will give rise to  emergent conserved quantities~\cite{Asmi_PRX} that will shatter the Hilbert space into disjoint sectors -- its own eigen-subspaces. All the Floquet states will hence respect the conservation law, and will hence be unable to exhibit Floquet thermalization.
In this procedure, the scars are {\it emergent}, and appears 
only due to the drive -- no conservation is needed to be imposed by hand, say, in form of projectors. Needless to stay, they affect the statistical mechanics by maintaining the conservation law. \\

\begin{figure}
\centering{
\includegraphics[width=0.85\linewidth]{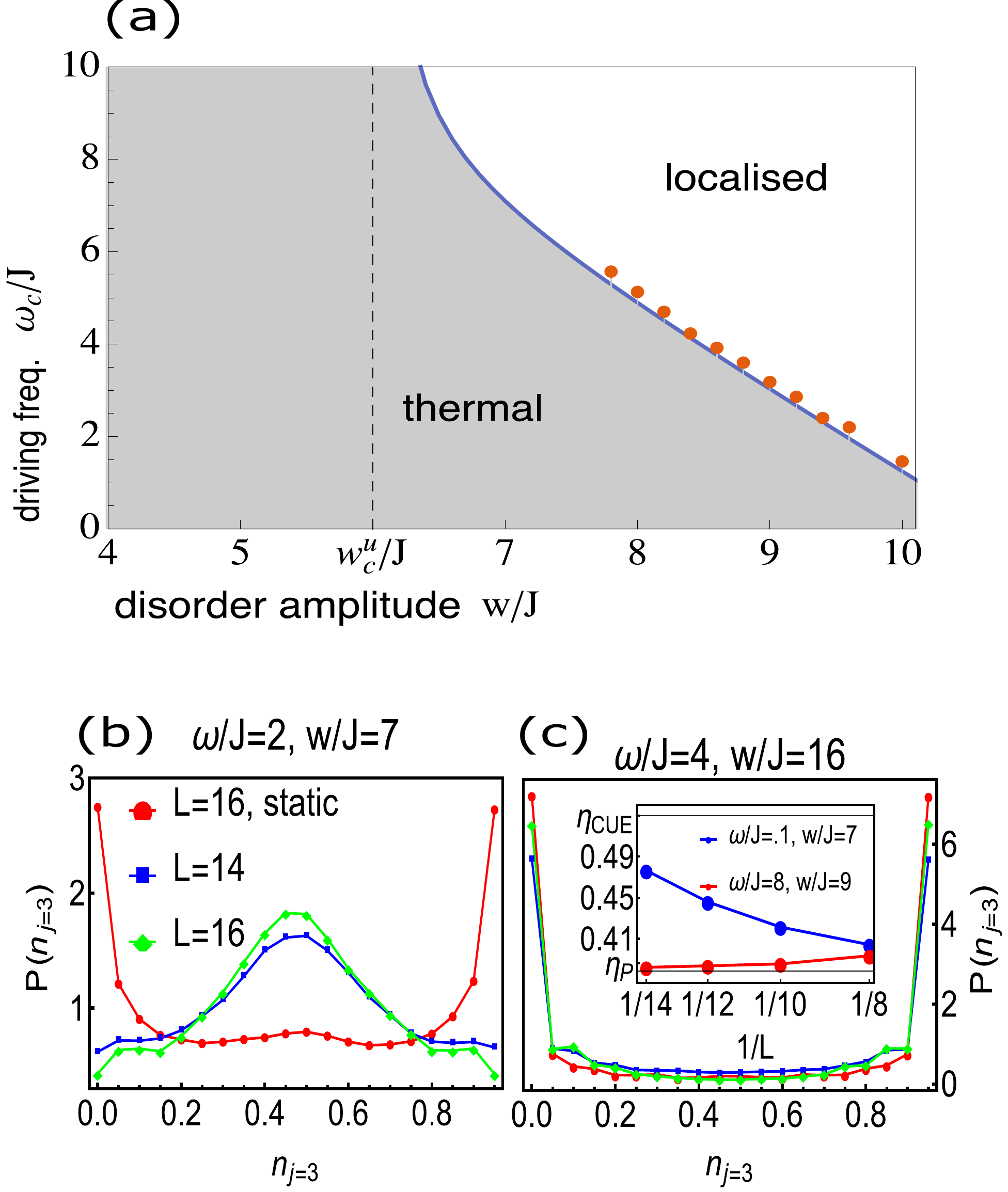}
}
\caption{
{\rm (a):}
Floquet-MBL phase diagram in the $w$ (disorder strength) and $\omega$
(drive frequency) plane. The shaded areas correspond
to Floquet thermalized (ergodic) phase. 
The red dots are obtained from finite-size studies
of the level statistics of the system. The disorder amplitude $w_{c}$
is the value below which the undriven system is delocalised in the
absence of driving. The blue line is a guide to the eye. \\
\noindent
{\rm (b)-(c):} 
Probability distribution of the eigenstate expectation values 
(EEVs) of the density at site $j=3$.
For low $\omega$
(b) Shows that for a low drive frequency $\omega$ and weaker disorder $w$, the Floquet eigenstates expectation values for the local density tends to get
more uniform over the Floquet spectrum (compared to the static EEV) with increasing system-size, indicating the approach to Floquet thermalization.
(c) Shows no such significant spreading for a shtonger disorder and higher drive frequency. Data is disorder averaged over $10^4$ (100) realisations for $L=14$ ($L=16$). Inset: Level statistics parameter versus inverse system size in the localised (bottom, red) delocalised (top, blue) phases. The parameter $\eta=\int ds\,s P(s)$ with $P(s)$ the probability distribution of the level statistics, taking the value $\eta_{P/CUE}$ in the localised/delocalised regime. Data averaged over $1000$ disorder realisations for $L=8, 10, 12$, $100$ realisations for $L=14$. (Taken from~\cite{AL_AD_RM_PRL_2015}).
\label{Flq_MBL_Fig}
}
\end{figure}
\section{Many-Body Localization Under Periodic Drive: Floquet MBL}
\subsection{Phenomenology}
Many body localization is the phenomenon where the eigenstates (including those with finite energy densities) of an interacting many-body system are found to be localized in an unentangled basis due to  disorder~\cite{Gornyi_MBL,Altshuler_MBL,Imbrie_PRL,Pal_Huse,Vadim_MBL,Abanin_lbits,Prosen_MBL,Kjall_PRL,Arti_MBL,MBL_Annl_der_Phys,Dima_RMP,Alet_MBL_Rev}. Technically, as shown in~\cite{Altshuler_MBL}, the 
localization is stable to all order in perturbation theory, the
perturbation being the interaction. So far no evidence of possible instabilities due to non-perturbative effects has been found in exact numerical results obtainable for small systems. The phenomenology of 
stable MBl can be rephrased in terms of an extensive local integrals of motions (LIOM) or $l-$bits~\cite{Imbrie_PRL,Abanin_lbits,Vadim_MBL}.\\

A major question was, what happens to an MBL system subjected to a periodic drive -- does the system absorbs heat and go to a Floquet thermzlized phase as happens in absence of disorder~\cite{LDM_PRE,Rigol_Infinite_T}, or does it persist in a localized phase? An intriguing picture comes out of the numerical studies~\cite{AL_AD_RM_PRL_2015,Abanin_MBL}. In summary, as shown 
in~\cite{AL_AD_RM_PRL_2015}, the result depends on whether the undriven system has a mobility edge: (a) if it does, the system always flows to the Floquet ETH scenario -- all the Floquet state looks locally infinite-temperature like, and (b) if it does not, 
the system might or might not heat up depending on 
the disorder strength $w$ and the drive frequency $\omega$
(Fig.~\ref{Flq_MBL_Fig}). The latter case, which supports an MBL-ergodic transition, was demonstrated 
for the following periodically driven MBL model.
\begin{equation}
	H_{0} =
		H_{hop}
		+\sum_{r=1}^{2} 
			V_{r}\sum_{i=1}^{L-1}n_{i}n_{i+r}
		+\sum_{i=1}^{L}U_{i}n_{i}\label{eq:H}
\end{equation}
where $H_{hop}=\left(-\frac{1}{2}J\sum_{i=1}^{L-1}\left(b_{i}^{\dagger}b_{i+1} + b_{i+1}^{\dagger}b_{i} + hc \right)\right)$ is a hopping operator, the $b$ are hard-core bosonic operators, $U_{i}$ an on-site random potential uniformly distributed between $-w$ and $+w$ and $H_{D}\left(t\right)$ a time-periodic hopping term
\begin{equation}
	H_{D}\left(t\right)=\delta\tilde{\delta}(t) H_{hop}
	\label{eq:periodic-hopping}
\end{equation}
with $\delta$ a dimensionless constant, $\tilde{\delta}(t)=-1(+1)$ in the first (second) half of each period $T=2\pi/\omega.$ 
The results for the Floquet thermalized and Floquet MBL states are shown and explained in Fig.~\ref{Flq_MBL_Fig}(b) and (c). These two frames corresponds to
two points on tow opposite sides of the phase boundary shown in frame (a) of the same Fig. From the physical point of view the scenario can be explained as follows. \\

\noi
In an undriven MBL, there is usually a classical potential-like (mutually commuting) part of the Hamiltonian (let us call it $H_x$), and there is a part that provides the quantum fluctuations (hopping/spin-flip etc). In case of Hamiltonian in Eq.~\ref{eq:periodic-hopping},  the density dependent part ($H - H_{hop}$) is the classical part, while $H_{hop}$ is the source of quantum fluctuations.
The classical part is trivially localized -- its eigenstates are unentangled, and can be expressed as a product of single-site states. Now existence of MBL (especially in absence of mobility edge) implies that in presence of the quantum fluctuations (let us call it hopping hence forth) the eigenstates are still localized: though might not be expressible as a product of single-site states, they are still expressible as a product of states defined over conglomerations ($l-$bits) of a few neighbouring sites, to a very good approximation. Here a few means there is distance-scale $\xi$: participation of spins at site $j$
to an $l-$bit $\tau_{i}$ localized at a given site $i,$ falls as $e^{-{|i-j|}/\xi}$. While there can be entanglement between the sites within an $l-$bit, different $l-$bits are practically unentangled with each other. Neglecting the cross-$l-$bit entanglements, we can consider the $l-$bits as mutually commuting LCQs - they commute with each other, and the common-basis that diagonalizes them also diagonalizes the Hamiltonian. \\

\noi
In this parlance, the question of Floquet heating in an MBL is tantamount to 
asking if the drive can hybridize and delocalize the $l-$bits to produce locally infinite-temperature like Floquet eigenstates. The analysis of the numerical results and the Floquet MBL-ergodic phase diagram (Fig.~\ref{Flq_MBL_Fig}) 
shows that at sufficiently high $\omega,$ when the drive is off-resonant with the local spectrum (roughly the spectrum of individual $l-$bits), there is no 
energy absorption in the local level, and the system does not heat up. This is
reflected in the fact that the Floquet eigenstates are still localized, and the drive just dresses up the $l-$bits~\cite{Keyserlingk_Sondhi_Vedika_Abs_Stabl_Flq}. 
These stable {\it Floquet $l-$bits are the emergent conserved quantities} that provides a Floquet MBL phase its stability. As in the case of static MBL, stability of the $l-$bits is based on the assumption that their exists a quasi-local unitary $U_{l}$ that diagonalizes the $\heff$ (or, in other words, assuming a finite depth unitary connects the localized basis and the Floquet eigenstates). The higher order processes (which makes $U_{l}$ more non-local, or increases its depth) that could hybridize the $l-$bits are suppressed due to the high frequency, as can be seen, e.g., from the Magnus expansion for the $\heff$ (Eq.\ref{Def:Magnus}). 
This, of course, is an {\it ex post facto} explanation, and could not be guessed
unless the convergence of the Maugnus expansion is proved. 
The suppression of heating needs a critical disorder strength, which monotonically 
decrease with increasing $\omega.$ As $\omega,$ is lowered, the higher order processes that hybridizes the $l-$bit becomes important, and Floquet thermalzation sets in.
\\

\noi
Strong Floquet drive can in general be used to tune the effective strength of the disorder, 
interactions, hopping in many-body Hamiltonian and hence can serve as a smooth handle for exploring
and controlling the non-equilibrium landscape of phases~\cite{Analabha_DL,Drive_Induced_MBL_Gill,Drive_Induced_MBL_Abanin}. 
\\

\subsection{An Example of a Floquet MBL Phase: The Discrete Time Crystal}

A zoo of Floquet MBL phases - both which break the symmetries of $H(t)$ and which do not, have been studied and classified (see, \cite{Keyserlingk_Sondhi_Flq_Phs_I,Keyserlingk_Sondhi_Flq_Phs_II} and references therein). Among those, some are absolutely stable under arbitrary small perturbations~\cite{Keyserlingk_Sondhi_Vedika_Abs_Stabl_Flq}. Here we will not go into the details of those classifications, but briefly discuss one interesting example of an absolutely stable Floquet MBL phase, namely, the discrete time crystal (DTC)~\cite{DTC_1_Vedika_ACL,DTC_2_Norm_Ashwin,Lukin_DTC,Monroe_DTC,DTC_4_Brief_History,DTC_5_Else_Norm_Rev}. For an alternative perspective of DTC see~\cite{DTC_3_Sacha,DTC_5_Sacha_Rev,Sacha_Book}.  

The possibility of breaking the continuous time-translation symmetry
of a time-independent many-body Hamiltonian in a
sense similar to the breaking of continuous translational symmetry in space by a crystal was considered by Wilczek~\cite{Wilczek}. It was soon showed by Watanabe and Oshikawa~\cite{Watanabe_Oshikawa} that such a ``time-crystal" phase is not possible in a quantum many-body system at equilibrium. However, it was subsequently shown, for
many-body Floquet Hamiltonian exhibiting discrete time translational symmetry: $H(t) = H(t+T),$ it is possible to have stable phases where this symmetry is broken by a stable state.
Such a discrete time-crystal breaks the discrete time-translation symmetry in its response in a ``crystalline" way, i.e., shows stable periodic but sub-harmonic response 
(e.g., a period doubled response) for special observables, starting from special initial states. The stability of the response to arbitrary but small perturbation in the 
time-lattice underlying the Hamiltonian, elevated DTC to the status of a non-equilibrium phase of matter. \\

For illustration,
let us consider a Floquet unitary defined over a drive period $T$ of the form
\beq
U_{0}(T) = Exp{\left[-i\pi \sum_{j}S_{i}^{y}\right]}
e^{-i H_{x}T}
\eeq 
\label{H_DTC_0}
\noi
which, applies on a set of spin-1/2 $S_{i}$s interacting via some Ising interactions in the $x-$direction given by $H_{x}(S_{i}^{x}),$ which is the classical part here. 
When applied for a period on any 
member of the $x-$basis (i.e., a simultaneous eigenstate of all $\sigma_{i}^{x}$), it will just flip all the spins and 
provide an overall dynamical phase due to $H_{x}.$ The 
original $x-$bais state just flips into its spin-flip partner state and no entanglement is produced. On the next cycle, the flipped spins will get a flip again, and the initial $x-$basis state will return to itself, modulo an overall dynamical phase.
This clearly means, a cat-like equal superposition of an $x-$basis states and its spin-flip partner will be a Floquet eigenstate, which will return to itself after every period. The system will hence be strongly non-thermal. 

However, if one introduces a small error/perturbation $\varepsilon$ in the period $\pi,$, say employ an unitary of the form
\beq
U_{\varepsilon}(T) = \exp{\left[-i(\pi +\varepsilon) \sum_{j}S_{i}^{y}\right]}e^{-i H_{x}T}, 
\eeq 
\label{H_DTC_eps}
\noi
then the action of $H_{x}$ is no longer ensured to produce an innocuous overall phase factor, but is potent of producing entanglement 
and delocalization over $x-$basis in every cycle, starting from the second one, even taking an 
$x-$basis state as the initial state. 
If $H_{x}$ contains generic (non-integrable) interactions, the system in general will heat up without bound
then~\cite{LDM_PRE,Alessio_Rigol_PRX}. However, if $H_{x}$
contains disorder, then this unbounded heating might get arrested, since it is then a candidate for forming a 
Floquet MBL~\cite{AL_AD_RM_PRL_2015, Abanin_MBL}. This stability 
has indeed been observed, both theoretically and experimentally
(see the references at in the beginning of the section).


\section{Experiments: A Brief Outline}
Floquet quantum matter can be realized in various setups, for example, in ultra-cold atoms in optical lattice~\cite{Bloch_Dalibard_RMP, Bloch_Gross_Sc_Rev_2017, PGE_Exp_Weld, Andre_RMP}, Rydberg systems~\cite{Scar_Lukin_Nature}, trapped ions~\cite{Blatt, Monroe_53_Qubit}, NMR~\cite{Rovny, Mahesh_Laflamme, Cappellaro},
NV centers~\cite{Lukin_DTC} and more. \\

Dynamical Many-body Freezing was demonstrated in an NMR setting for an integrable Ising chain~\cite{Mahesh_DMF_2014}.
There, the average magnetization vs drive frequency, including the freezing peaks were observed, 
in surprisingly good agreement with the analytical formula (Eq.~\ref{Q_Analytic}) derived in~\cite{AD-DMF}. \\

\noi
A clear experimental demonstration of the periodic Gibbs' ensemble (PGE), the Floquet thermalized ensemble, and crossing over from the latter to the former by tuning an integrability breaking interaction has been reported in Ref.~\cite{PGE_Exp_Weld}. There, in a periodically driven BEC, the fraction of atoms
in the ground state (bose condensate) was used to quantify the extent of thermalization. The non-interacting as well the interacting versions are studied, and the non-interacting case, the results were consistent with the PGE predictions~\cite{PGE}.
Eventually, at late times, the system crosses over to the locally infinite-temperature like Floquet ETH  phase~\cite{LDM_PRE,Alessio_Rigol_PRX}.
\\

\noi
Interesting demonstration of the prethermal regime (exponential suppression of heating with increasing drive
frequency)~\cite{Abdal_Vedika_Bloch} and persistence of
quasi-conservation laws~\cite{PGE} past the expected prethermalization time~\cite{Cappellaro} has been reported more recently.\\

\noi
The existence of Floquet MBL phases~\cite{AL_AD_RM_PRL_2015,Abanin_MBL}
are also interesting numerical studies that confirmed the stability of the Floquet
MBL phase~\cite{Bordia_Bloch_Ulrich_Flq_MBL,Bordia_Knap_Bloch}, especially, in the context of the discrete time crystals~\cite{Lukin_DTC,Monroe_DTC,Rovny,Sreejith_Mahesh}.
\\

\section{Summary and Outlook}
In this review we broadly explore the structure of the statistical mechanics
of periodically driven closed quantum matter or Floquet matter. The focus was on
various ensembles and the local conserved quantities (LCQ) that characterize them. In addition to the exact periodically conserved quantities that appears, say, due to integrability of the system, the structure of the statistical mechanics is shaped also by conservation laws and symmetries 
that emerges {\it due to the drive} (not present in the undriven system).
In clean, interacting cases, they prevent unbounded heating or the Floquet thermalization. In MBL systems, the Floquet $l-$bits emerges as the LCQs, and provide absolute stability to various interesting Floquet states, including the discrete time crystals. \\

Identifying the emergent symmetries, constraints and conservation laws in interacting Floquet system, and tailoring them to engineer novel Floquet phases
of matter is a potentially rich future direction. The absolute stability of such phases in the thermodynamic limit is a harder open question which requires 
analytical approaches involving (most likely divergent) perturbation series and beyond. However, an intermediate scale of about 100 atoms/spins are probably simulatable in near future in quantum computers (see, e.g.~\cite{Sycamore_Vedika}). 

During last few decades, experimental developments in various setups including ultra cold atoms in optical lattice, ion traps, NV centers and NMR simulators enabled realization of Floquet quantum matter in a controlled way in the laboratories.
This brings up the importance of statistical characterization of the 
intermediate-size ($\sim 100$ atoms/spins) samples of Floquet matter. 

Non-equilibrium fluctuations relations\cite{Fluctuation_Relations_Hanggi,Jarzynski}, statistics of large deviations~\cite{Large_Deviation_Hugo_Touchette}, are some directions in which investigating Floquet matter~\cite{Kris_AD_AD_Flq_Work_Distrb} could be interesting. \\

\noindent
{\textbf{\textit{Acknowledgements}}}: 
The authors thank S.Bhattacharyya, S. Dasgupta, A. Dutta, S. S. Hegde, H. Katiyar, A. Lazarides, T. S. Mahesh, R. Moessner, 
A. Roy, D. Sen and K. Sengupta for collaborations at various stages of the developments. We thank D. Sen for a careful 
reading of the MS and his comments on it.

\bibliographystyle{apsrev4-1}
\bibliography{Bib_Onset_Scar_QPT}

\end{document}